\documentclass[journal]{ieeetmlcn} %IEEETran
\usepackage{cite}
\usepackage{amsmath,amssymb,amsfonts}
\usepackage{algorithm}
\usepackage{algpseudocode}
\usepackage{graphicx, color}
\usepackage{textcomp}
\usepackage{xcolor}
\usepackage{comment}
\usepackage{url}

\usepackage{hyperref}
\hypersetup{hidelinks}

\makeatletter
\renewcommand{\p@subsection}{\thesection-} % produces II-A in \ref/\autoref

\def\BibTeX{{\rm B\kern-.05em{\sc i\kern-.025em b}\kern-.08em
    T\kern-.1667em\lower.7ex\hbox{E}\kern-.125emX}}

\AtBeginDocument{\definecolor{tmlcncolor}{cmyk}{0.93,0.59,0.15,0.02}\definecolor{NavyBlue}{RGB}{0,86,125}}

\def\OJlogo{\vspace{-4pt}\includegraphics[height=18pt]{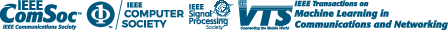}}
\def\seclogo{\vspace{10pt}\includegraphics[height=18pt]{new_logo}}

\usepackage{float}
\usepackage[caption=false]{subfig}

\begin{document}
\receiveddate{XX Month, XXXX}
\reviseddate{XX Month, XXXX}
\accepteddate{XX Month, XXXX}
\publisheddate{XX Month, XXXX}
\currentdate{XX Month, XXXX}
%\doiinfo{TMLCN.2022.1234567}

\markboth{}{Author {et al.}}

\title{Efficient Channel Autoencoders for Wideband Communications leveraging Walsh-Hadamard interleaving}
\author{Cel Thys, Rodney Martinez Alonso, Sofie Pollin}
\affil{WaveCoRE, Department of Electrical Engineering (ESAT), Katholieke Universiteit Leuven, Belgium}
\corresp{Corresponding author: C. Thys (email: cel.thys@kuleuven.be).}
\authornote{This work was supported by the European Union's Horizon 2020 research and innovation programme under grant agreement No 964246 (HERMES). This work only reflects the author's view, and the European Commission is not responsible for any use that may be made of the information it contains. The work of Rodney Martinez Alonso was supported by the Research Foundation–Flanders (FWO) under Grant 1211926N.}

\begin{abstract}
This paper investigates how end-to-end (E2E) channel autoencoders (AEs) can achieve energy-efficient wideband communications by leveraging Walsh-Hadamard (WH) interleaved converters. WH interleaving enables high sampling rate analog-digital conversion with reduced power consumption using an analog WH transformation. We demonstrate that E2E-trained neural coded modulation can transparently adapt to the WH-transceiver hardware without requiring algorithmic redesign. Focusing on the short block length regime, we train WH-domain AEs and benchmark them against standard neural and conventional baselines, including 5G Polar codes. We quantify the \emph{system-level} energy tradeoffs among baseband compute, channel signal-to-noise ratio (SNR), and analog converter power. Our analysis shows that the proposed WH-AE system can approach conventional Polar code SNR performance within $0.14$\,dB while consuming comparable or lower system power. Compared to the best neural baseline, WH-AE achieves, on average, 29\% higher energy efficiency (in bit/J) for the same reliability. These findings establish WH-domain learning as a viable path to energy-efficient, high-throughput wideband communications by explicitly balancing compute complexity, SNR, and analog power consumption.
\end{abstract}


\maketitle

\section{Introduction}
Since the advent of communication technology, communication engineers have worked to increase the energy efficiency and throughput of these systems. Historically, this has been motivated by the steady growth in network traffic and by power-consumption constraints in mobile devices. More recent market drivers include high infrastructure energy costs and emerging high-throughput applications, such as extended reality or artificial intelligence (AI) training~\cite{jornet_evolution_2024}, consistently fueling the need for higher throughput and higher energy efficiency. 

To increase throughput means either improving spectral efficiency or increasing bandwidth. Approaches to increase bandwidth for wideband communication links typically require more energy-intensive hardware due to the higher sampling rate in the conversion hardware. Higher spectral efficiency, on the other hand, comes with a compute or energy cost in the baseband. This paper proposes an AI-based exploration framework to optimize the tradeoff between throughput and energy efficiency. We present a Walsh-Hadamard domain modulation and coding scheme that leverages energy-efficient hardware to improve that tradeoff.

Recent examples of wideband communication systems include the allocation of C-band and mmWave spectrum in 5G, as well as research on spectrum bands such as FR3~\cite{cui_6g_2025}, sub-terahertz~\cite{chakraborty_twirld_2024}, and free space optical~\cite{elfikky_symbol_2024}. By widening the bandwidth, simple modulation and coding schemes can be employed without resorting to complex baseband hardware to improve throughput. 
Although the new spectrum bands all present unique challenges related to hardware impairments and different propagation behaviour~\cite{jornet_evolution_2024,cui_6g_2025}, one common limitation is the power consumption of the domain converter hardware: Analog-to-Digital and Digital-to-Analog Converters (ADCs and DACs)~\cite{adc_survey}.

Converter interleaving~\cite{balasubramanian_systematic_2011, schmidt_data_2020,buchwald_high-speed_2016} is a technique used to implement energy-efficient, high-rate ADCs and DACs for wideband communications. The technique involves parallelizing the converter into a converter array, where each sub-converter can operate independently at a significantly lower sampling rate. Reducing the sampling rate decreases power consumption and simplifies the sub-converter design~\cite{manganaro_introduction_2022, buchwald_high-speed_2016}. One common interleaving approach is time interleaving (TI), in which the sub-converters operate on successive sub-samples of the original signal. This sequential operation of the sub-converters requires a carefully designed sample-and-hold (S\&H) frontend~\cite{buchwald_highspeed_2016} with high time resolution.

To relax the time-resolution requirements of interleaved converters, one can use frequency-domain or Walsh-Hadamard (WH) domain interleaving. In frequency interleaved converters~\cite{schmidt_data_2020, jornet_evolution_2024}, each sub-converter is dedicated to one specific subband of the original signal, where the subbands are created using an analog filter bank that reduces at the same time the required time resolution in each sub-branch. WH interleaving is a more energy-efficient interleaving technique where the parallel converters operate concurrently in the WH domain~\cite{dehos_d-band_2022, bouassida_concurrent_2016, ferrer_walsh-based_2023, ferrer_experimental_2025, fellmann_block-based_2023, thys_walsh-domain_2024}. In replacing the analog filter bank used in frequency interleaving by mixing with digital Walsh-Hadamard codes, the complexity and power consumption are further reduced. Thus, among all interleaved converter approaches, WH interleaving promises the highest energy efficiency for wideband communications. 

An open problem for WH interleaving is how to design the modulation and coding system for this novel converter architecture, while a systematic end-to-end energy-efficiency analysis that accounts for both analog and digital processing is not yet available in the state of the art. AI and machine learning are now becoming widely used to improve communication system performance~\cite{hoydis_2021_6gainative}, and can readily be adapted to the WH-domain modulation and coding problem. The critical insight for leveraging AI in the digital baseband is that one can model the communication system as an autoencoder~\cite{oshea_introduction_2017}, comprising an encoder at the transmit side, a decoder at the receive side, and a channel model. A growing body of literature has since extended the channel autoencoder idea to different modulation schemes~\cite{felix_ofdm-autoencoder_2018, jiang_turbo_2019}, new spectrum bands~\cite{chakraborty_twirld_2024,elfikky_symbol_2024} or hardware constraints~\cite{ait_aoudia_waveform_2022,honkala_deeprx_2021,marasinghe_waveform_2025,dorner_learning_2023,wiesmayr_design_2024}. The main benefits of this approach are: all parts of the system can be optimised jointly to operate as close as possible to the fundamental limits for finite block length communication~\cite{polyanskiy_channel_2010, cammerer_trainable_2020, rajapaksha_low_2020, hesham_coding_2023, dorner_learning_2023}, and the learning system can be easily adapted to support novel modulation schemes, channel conditions or hardware constraints~\cite{hoydis_2021_6gainative}.

\begin{figure*}[tb]
\centering
\includegraphics[width=0.9\textwidth, keepaspectratio]{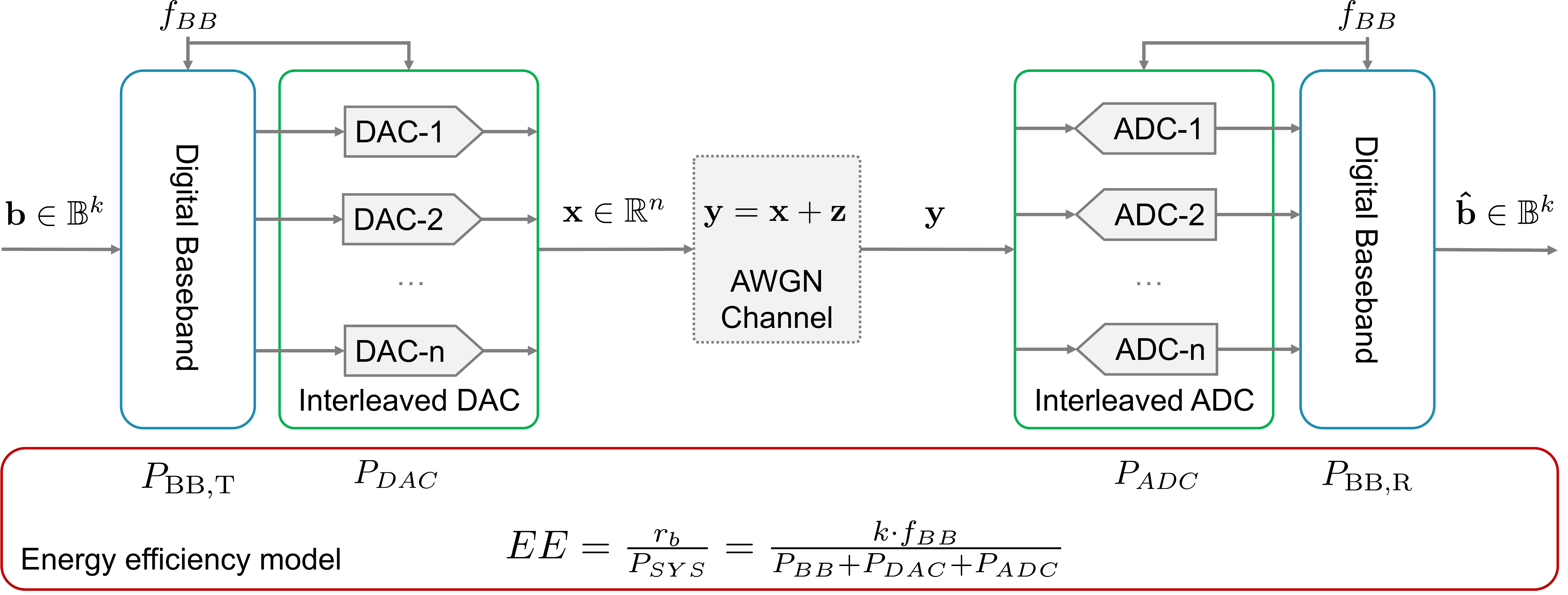}
\caption{Proposed wideband transceiver architecture including high-rate interleaved DAC \& ADC and intelligent AI baseband processing. The energy efficiency model takes into account the interleaving converter power, the baseband power, the baseband throughput and the channel SNR.}
\label{fig:walsh_transceiver}
\end{figure*}

Our work aims to design an energy-efficient, wide-bandwidth communication system by combining the two aforementioned approaches: WH interleaved converters and machine learning for finite block length communications. More specifically, our contributions are the following:
\begin{itemize}
\item We propose the Walsh-Hadamard interleaved autoencoder, an energy-efficient communication system based on low-power data conversion and intelligent baseband design, that operates reliably close to the limits of finite block length coding.
\item We develop a detailed power consumption model for the AI baseband, which, combined with interleaved converter power consumption, allows us to assess the system-level energy efficiency.
\item We analyze the complexity-performance tradeoff of the proposed transceiver architecture and conventional baselines, balancing communication reliability and energy efficiency.
\end{itemize}

The outline of this manuscript is as follows. First, Section~\ref{sec:problem_formulation} introduces the proposed transceiver architecture, including the general system model for interleaved wideband communications and the details of WH-domain interleaved data conversion. Second, we build on this high-bandwidth transceiver model by introducing our proposed autoencoder for short block length modulation and coding in Section~\ref{sec:learned_modulation}. Key here is also the tradeoff between compute energy and coding performance, which motivates our total energy and performance model in Section~\ref{sec:powermodel}. The results in Section~\ref{sec:results} enable a detailed analysis of this tradeoff, which supports the conclusion drawn in Section~\ref{sec:conclusion}.

\textit{Notations}: Italic $x$ is used for a random variable, bold $\boldsymbol{x}$ represents a vector, and subscript w is used for the WH transform of the corresponding variable: $\boldsymbol{x}_w = \mathcal{W}(\boldsymbol{x})$. $\mathbb{E}_x\{y\}$ is the statistical expectation of variable $y$ with respect to all realizations of variable $x$. The Gaussian distribution is written $\mathcal{N}(\mu, \sigma)$, with mean $\mu$ and variance $\sigma^2$, while $\mathcal{U}(S)$ represents the uniform distribution over the set or domain $S$. The set of real numbers is denoted $\mathbb{R}$, the binary numbers $\mathbb{B}$, and the set of length-$k$ binary vectors is written $\mathbb{B}^k$. We use $\log_2$ to represent the binary logarithm and $\log_{10}$ for the base 10 logarithm. Finally, matrices are written in uppercase, and $ C = A \otimes B $ represents the Kronecker product of two matrices.

\section{Wideband transceiver model} \label{sec:problem_formulation}
The proposed wideband transceiver architecture for high-rate, high-efficiency communications, as shown in Fig.~\ref{fig:walsh_transceiver}, consists of wideband interleaved converters and an intelligent baseband that drives them in an energy-efficient manner. The general system model is introduced in \autoref{subsec:system_model}, followed by a detailed description of interleaved conversion and signal representation in the WH domain in \autoref{subsec:walsh_theory}.

\subsection{System model}\label{subsec:system_model}
The considered baseband communication system consists of a transmitter, a communication channel model, and a matching receiver. Both the transmitter and the receiver consist of a digital baseband and an interleaved domain converter (DAC and ADC, respectively). Interleaving is the key technique that enables high efficiency wideband communications, by allowing each sub-converter to operate at the baseband rate
\begin{equation}
f_{\text{BB}} = \frac{f_s}{n} = \frac{2 B}{n}, 
\end{equation}
where $B$ is the channel bandwidth and $f_s=2 B$ the equivalent sampling rate. Due to interleaving, the baseband processes $n$ real-valued symbols at a rate of $f_{\text{BB}}$ (for details see~\ref{subsec:walsh_theory}).

The transmitter performs joint coding and modulation with information rate $R=\frac{k}{n}$. Here, $k$ is the number of information bits encoded in one modulation symbol and $n$ is the number of real-valued transmissions ('channel uses') per symbol. More specifically, the transmitter encodes input bit vectors $ \boldsymbol{b} \in \mathbb{B}^k$ into modulation symbols $\boldsymbol{x} \in \mathbb{R}^{n}$, which are sent for transmission over the point-to-point Additive White Gaussian Noise (AWGN) channel:
\begin{IEEEeqnarray}{rCl}
    \boldsymbol{x} & = & f (\boldsymbol{b}), \\
    \boldsymbol{y} & = & \boldsymbol{x} + \boldsymbol{z}. \label{eq:awgn_real}
\end{IEEEeqnarray}

The receiver decodes the channel responses $\boldsymbol{y} \in \mathbb{R}^{n}$ to bit probability estimates $\hat{\boldsymbol{p}} \in \mathbb{R}^k_{+}$. These probabilities can be hard-decision decoded to yield bit estimates $\hat{\boldsymbol{b}} \in \mathbb{B}^k$:
\begin{IEEEeqnarray}{rCl}
    \hat{\boldsymbol{p}}(\boldsymbol{b}|\boldsymbol{y}) & = & g (\boldsymbol{y}), \\
    \hat{\boldsymbol{b}} & = & \hat{\boldsymbol{p}}(\boldsymbol{b}|\boldsymbol{y}) > 0.5. \label{eq:hard_decision}
\end{IEEEeqnarray}
Here, $f$ and $g$ represent the digital baseband encoding and decoding, which can be conventional or trainable nonlinear operators. The vector $\hat{\boldsymbol{p}}(\boldsymbol{b}|\boldsymbol{y})$ contains the estimated posterior bit probabilities that are used during system training, while \eqref{eq:hard_decision} describes the hard-decision decoding process used during evaluation or deployment to get received bit estimates.

Equation~\eqref{eq:awgn_real} models the discrete-time AWGN channel with noise $\boldsymbol{z}$, a vector of $n$ i.i.d. real Gaussian random variables with mean $0$ and standard deviation $\sigma_z$. The channel input $\boldsymbol{x}$ is constrained to unit average power for fair comparison with conventional modulation:
\begin{IEEEeqnarray}{c}
    \boldsymbol{z} \sim \mathcal{N}(0, \sigma_z^2 \boldsymbol{I}_n), \\
    \mathbb{E}_{\boldsymbol{b}} \left\{ \|\boldsymbol{x}_i\|^2 \right\} = 1. \label{eq:power_constraint}
\end{IEEEeqnarray}
By normalizing the codeword power, the channel SNR $\gamma$ is controlled by adjusting the noise standard deviation $\sigma_z$: 
\begin{equation}
    \gamma = \frac{1}{\sigma_z^2}. \label{eq:snr_conversion}
\end{equation}

In the evaluation in Section~\ref{sec:results}, the autoencoder system will be compared to the Shannon capacity for discrete-time real AWGN channels with unit average power constraint:
\begin{equation}
    C(\gamma) = \frac{1}{2}\log_2(1+\gamma) = \frac{1}{2}\log_2\left(1+\frac{1}{\sigma_z^2}\right). \label{eq:shannon}
\end{equation}

\begin{figure*}[tbp]
%\centering
%\begin{subfigure}[t]{0.46\linewidth}
%\centering
\subfloat[\label{fig:walshtime}]{
\includegraphics[width=0.46\linewidth, height=0.5\linewidth,keepaspectratio]{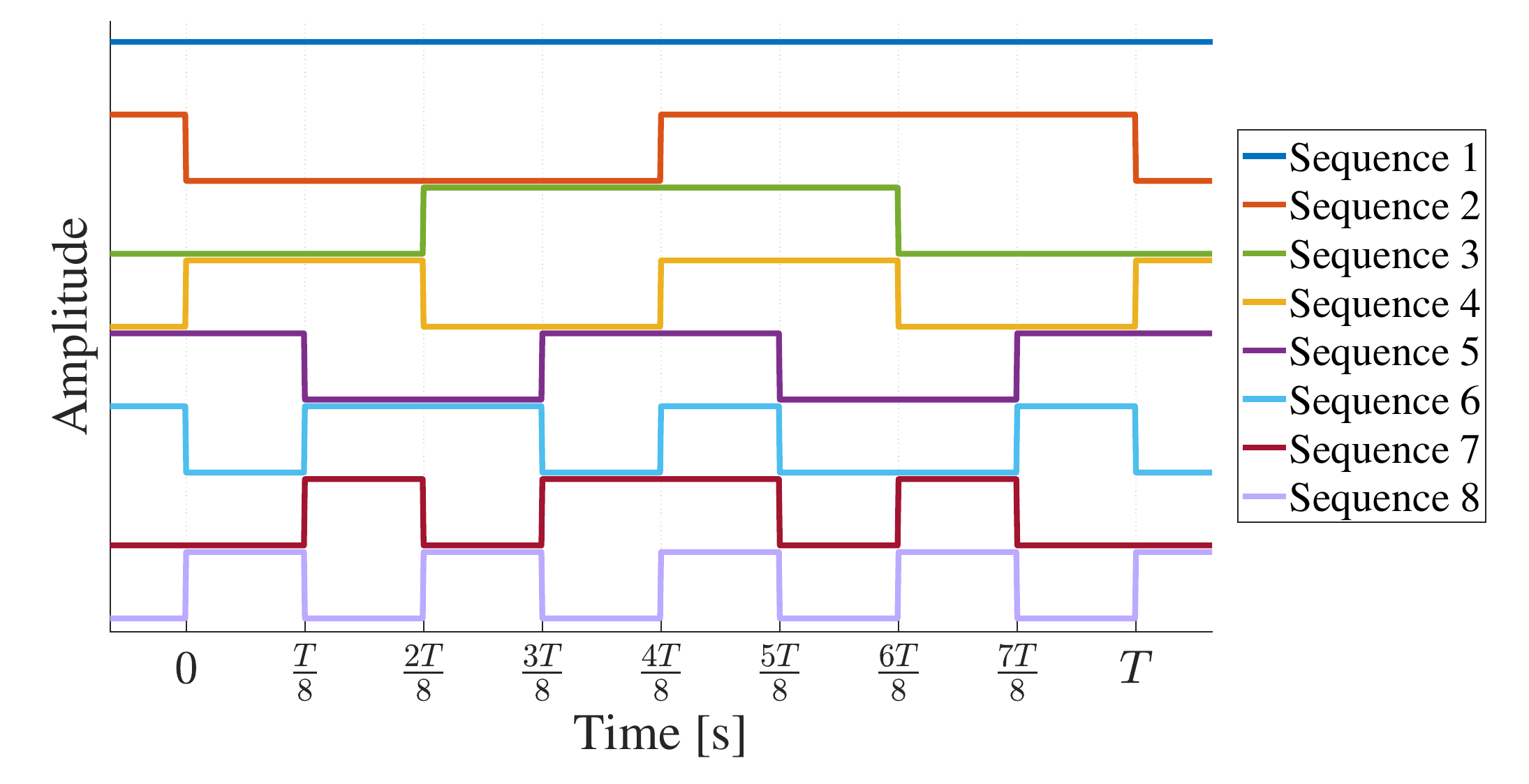}}
%\caption{}
%\label{fig:walshtime}
%\end{subfigure}
%\begin{subfigure}[t]{0.53\linewidth}
%\centering
\subfloat[\label{fig:walshfreq}]{
\includegraphics[width=0.53\linewidth, height=0.5\linewidth,keepaspectratio]{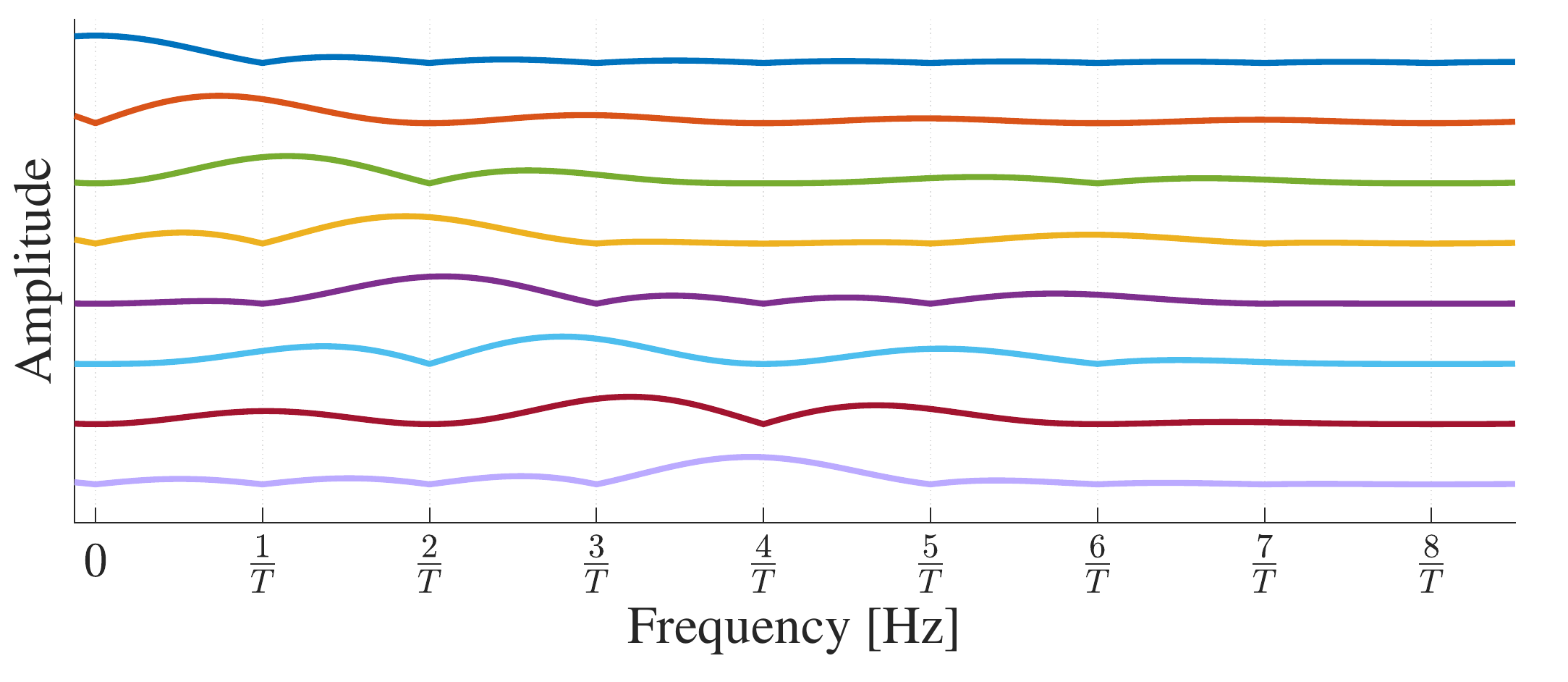}}
%\caption{}
%\label{fig:walshfreq}
%\end{subfigure}
\caption{Example of the first 8 sequency-ordered Walsh-Hadamard basis functions in (a) time and (b) frequency domain.} %The frequency spectra are symmetrical because the functions are real-valued. } % note: I also have a version with periodic Walsh functions and a discrete spectrum: walsh_freq.eps
\label{fig:walshbasis}
%\vspace{-3mm} % to reduce white space after figure
\end{figure*}

In the proposed practical system, we perform modulation and coding with a finite block length $n$, and thus we define the block-error-rate (BLER):
\begin{equation}
    P_e = P(\hat{\boldsymbol{b}} \neq \boldsymbol{b}).
\end{equation}
In this regime, the maximum achievable rate (in bits-per-channel-use) is bounded by the Shannon limit: $ R^* \leq C$ and depends not only on the SNR $\gamma$ but also on the block length $n$ and the BLER $P_e$ \cite[Eqn. (291)-(295)]{polyanskiy_channel_2010}:
\begin{IEEEeqnarray}{c}
    R^*(\gamma, P_e, n) = C(\gamma) - \sqrt{\tfrac{V(\gamma)}{n}} Q^{-1} (P_e) + \tfrac{3\log(n)}{2n}, \label{eq:rate_bound} \\
    V(\gamma) = \frac{\gamma (\gamma + 2)}{2 (\gamma+1)^2} \log_2^2(e).
\end{IEEEeqnarray}
It is clear from this equation that in the finite block length regime, the achievable rate for a given SNR $\gamma$ is strictly below the Shannon capacity, with an offset proportional to $Q^{-1} (P_e)$ with a SNR-dependent scaling factor $V(\gamma)$. Conversely, transmissions at a given rate $R$ require an SNR higher than the ideal Shannon SNR, with the offset depending on $n$ and the target BLER.

While \eqref{eq:rate_bound} provides an upper bound on the achievable rate in the finite block length regime, the key questions are how to design encoding and decoding schemes that achieve it, and how the proposed method compares with known schemes such as Polar codes and neural codes. In our approach, we first fix a BLER target of $P_e=0.1\%=10^{-3}$, similar to 5G NR Polar code design~\cite{bioglio_design_2021}, and then measure the error correction performance using the threshold SNR required for each modulation and coding scheme to achieve the BLER target.

More specifically, for a fixed $n$ and the fixed BLER target, we will define the threshold SNR as follows:
\begin{equation}
\gamma_{\text{th}} = \min {\gamma} \textrm{ where } P_e(f,g,\gamma) \leq 10^{-3}.\label{eq:threshold_snr}
\end{equation}
Here, the achieved BLER $P_e(f,g,\gamma)$ depends on the channel SNR $\gamma$, but also on the baseband encoding and decoding functions $f,g$. These functions can implement conventional Polar or neural autoencoder-based coding and modulation.

\subsection{Orthogonal Walsh-Hadamard domain conversion}\label{subsec:walsh_theory}
The system model in \autoref{subsec:system_model} is generic and does not describe the interleaved conversion method or the baseband implementation. In this section, we outline our proposed Walsh-Hadamard domain transceiver architecture, which combines energy-efficient WH-interleaved conversion and neural autoencoder baseband processing.

Figure~\ref{fig:walsh_transceiver} shows our proposed architecture for energy-efficient wideband communication systems based on interleaved domain conversion. Whereas conventional wideband converters operate in time-interleaved mode, the proposed architecture employs Walsh-Hadamard domain interleaving.

\textit{\textbf{Orthogonal Walsh-Hadamard transformation}:} The Walsh-Hadamard basis is a complete set of orthogonal functions that can be used to represent signals in a compact way, by using the inner product of the signal with the basis~\cite{harmuth_applications_1969}. The basis functions $\phi_{W,i},i=1,\dots,N$ are defined using the Hadamard matrix $H_N$ where $N$ is a power of 2:
\begin{equation}
    H_N = 
    \begin{cases}
        \begin{pmatrix}
            1 & 1 \\
            1 & -1
        \end{pmatrix}, \quad N = 2, \\
        H_{N/2} \otimes \begin{pmatrix}
            1 & 1 \\
            1 & -1
        \end{pmatrix}, \quad N > 2
    \end{cases}
\end{equation}
The Hadamard matrix is thus a square matrix of size $N \times N$ with entries of either $1$ or $-1$ and orthogonal rows. The Walsh matrix $W_N$ is a permutation of the Hadamard matrix where the rows are ordered in ascending sequency order, with sequency defined as the number of sign changes in a row~\cite{fino_unified_1976,fellmann_block-based_2023}. As an example, we list below the Hadamard and Walsh matrices for $N=4$:

\begin{equation}
    H_4 = \begin{pmatrix}
        1 & 1 & 1 & 1 \\
        1 & -1 & 1 & -1 \\
        1 & 1 & -1 & -1 \\
        1 & -1 & -1 & 1
    \end{pmatrix},
    W_4 = \begin{pmatrix}
        1 & 1 & 1 & 1 \\
        1 & 1 & -1 & -1 \\
        1 & -1 & -1 & 1 \\
        1 & -1 & 1 & -1
    \end{pmatrix}.
\end{equation}
In a similar way, the time- or frequency-domain orthogonal basis can be defined using the identity and the complex Fourier matrix, respectively. The structure of each basis matrix also reveals the main drawbacks of each approach: the time domain identity matrix is sparse, which reflects time-interleaving converters that require careful time synchronization\cite{buchwald_highspeed_2016}; the Fourier matrix is built using complex exponentials, reflecting the higher complexity of frequency interleaving compared to digital Walsh-domain interleaving, which only utilises sequences of $\pm1$. 

To focus on the discrete-time application of the Walsh-Hadamard transform, we define the Discrete Walsh-Hadamard Transform (DWHT). Let $\boldsymbol{x}$ be a vector of discrete samples of the signal $x[k]$, then one calculates the DWHT of $\boldsymbol{x}$ by multiplying the discrete time signal vector by $W_N$: 
\begin{align}
\boldsymbol{x} & = [x[0], \dots, x[N]],\\
\boldsymbol{x}_w & = W_N \cdot \boldsymbol{x} = \mathcal{W}(\boldsymbol{x}).
\end{align}
In this notation, we have limited the length of the discrete time signal $\boldsymbol{x}$ to $N$ samples, which is the size of the Walsh matrix. Vice versa, the notation $\mathcal{W}(\boldsymbol{x})$ assumes the input length $N$ is a power of 2.

An interpretation of the DWHT is to consider it a projection of the discrete time signal $x[k]$ onto a basis of oscillating functions $\phi_{W, i}[k]$ defined by the rows of the Walsh matrix (with faster oscillations for increasing row number $i$), similar to a Discrete Fourier Transform (DFT). This projection domain is called the Walsh or sequency domain, where the row number $i$ is the \emph{sequency} of the corresponding basis function. Figure~\ref{fig:walshbasis} shows the Walsh-Hadamard basis functions of order $N=8$ with ascending sequency, in time and frequency domains. The time domain functions are rectangular pulses, and the frequency domain functions are real-valued, with single-sided spectra that act as frequency-selective filters. The passband of the frequency-selective filter shifts with increasing sequency $i$.

Similar to the DFT, the DWHT transform can be efficiently implemented in a butterfly structure~\cite{fino_unified_1976}. For the DFT, this process is called the Fast Fourier Transform (FFT), while for the Walsh transformation, it is called the Fast Walsh-Hadamard Transform (FWHT). While the Fourier transformation uses complex exponentials, both the FWHT and its inverse (the IFWHT) can be implemented solely using addition and subtraction of the real-valued input. In summary, the FWHT can be considered a low complexity, real-valued alternative to the FFT.

To illustrate how the digital Walsh-Hadamard transform can be implemented in analog-digital domain converters, we provide a brief description of the continuous-time Walsh–Hadamard transform here, as implemented in physical circuits~\cite{dehos_d-band_2022, bouassida_concurrent_2016, ferrer_walsh-based_2023, ferrer_experimental_2025}. The continous-time basis function $\phi_{W,i}(t)$ is derived from the $i$-th row of the Walsh matrix $W_N$, by dividing the symbol interval of length $T$ into $N$ equal parts and assigning subsequent values of the $i$-th row to each part: 

\begin{equation}
\phi_{W,i}(t) = W_N(i,j) \text{ for } t \in \left[ \frac{(j-1)T}{N}, \frac{jT}{N} \right).
\end{equation}

The $N$-th order continous Walsh-Hadamard transform, for a continous signal $x(t)$ defined on the time interval $[0,T]$, is defined as:
\begin{IEEEeqnarray}{rCl}
x(t) & = & \sum_{i=1}^{N} \left< x(t), \phi_{W,i}(t) \right> \cdot \phi_{W,i}(t),\\
x(t) & = & \sum_{i=1}^N x_{W}[i] \cdot \phi_{W,i}(t). \label{eq:reconstruction}
\end{IEEEeqnarray}
The inner products $x_{W}[i]$ are also called Walsh-Hadamard coefficients, as they represent the projection of the signal $x(t)$ onto the basis. A Walsh-domain interleaved ADC samples in each of the interleaved branches one of these coefficients by mixing the input signal with the relevant basis:
\begin{equation}
x_W[i] = \int_{0}^{T} x(t) \cdot \phi_{W,i}(t) dt, \quad i=1,\dots,N.
\end{equation}
The Walsh-domain interleaved DAC performs the inverse reconstruction from the digital WH-domain coefficients to an analog time domain output signal, by mixing the sub-converter outputs with the basis functions $\phi_{W,i}$ and summing the result. The reconstruction of the original signal $x(t)$ can be modeled by means of Eqn.~\ref{eq:reconstruction} or its digital version (the IFWHT): $\boldsymbol{x} = \mathcal{W}^{-1}(\boldsymbol{x}_w)$. 

\begin{figure}[htbp]
\subfloat[\label{fig:walshadc}]{
\includegraphics[width=0.44\linewidth, height=0.5\linewidth,keepaspectratio]{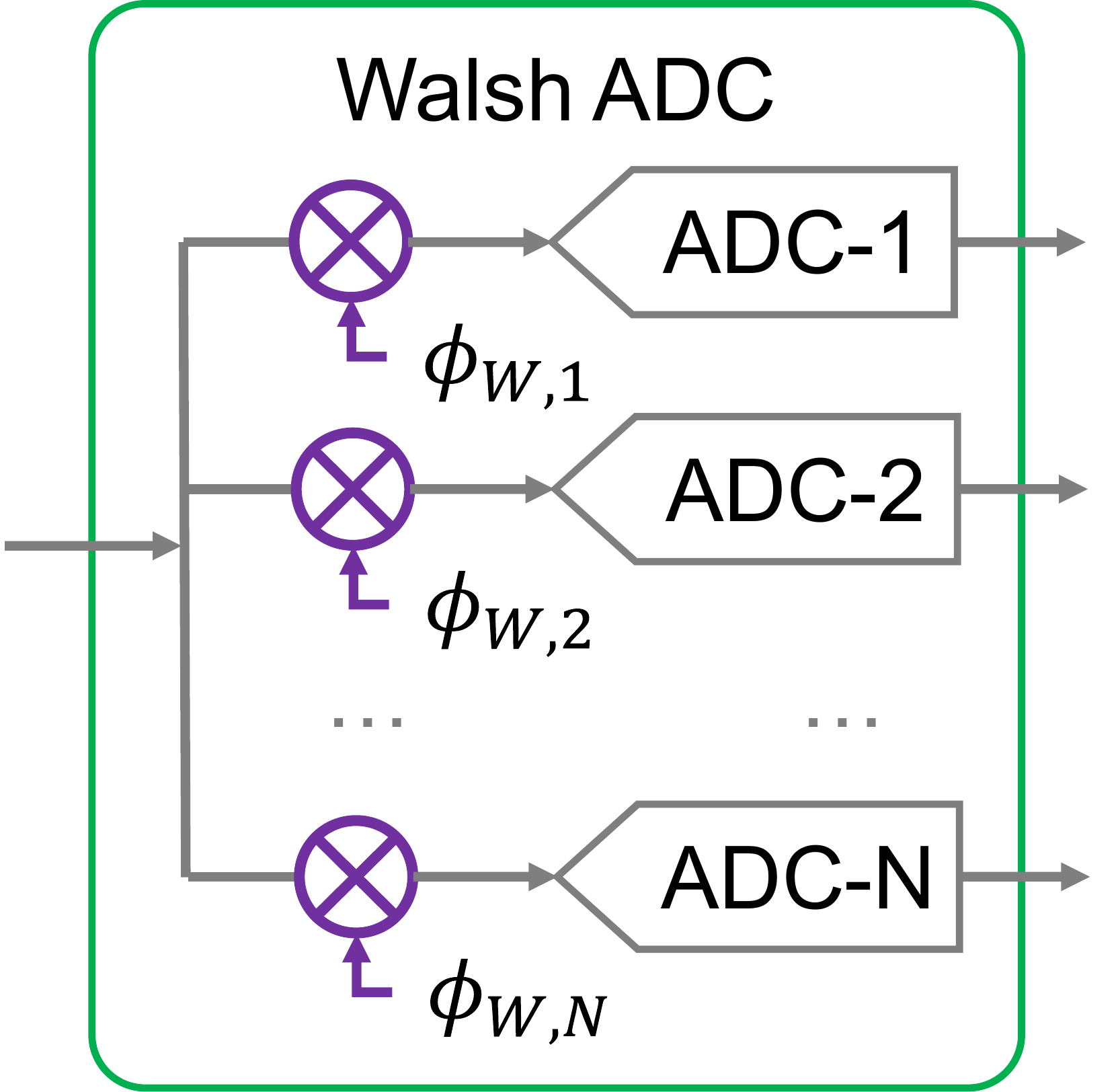}}
\subfloat[\label{fig:walshdac}]{
\includegraphics[width=0.47\linewidth, height=0.5\linewidth,keepaspectratio]{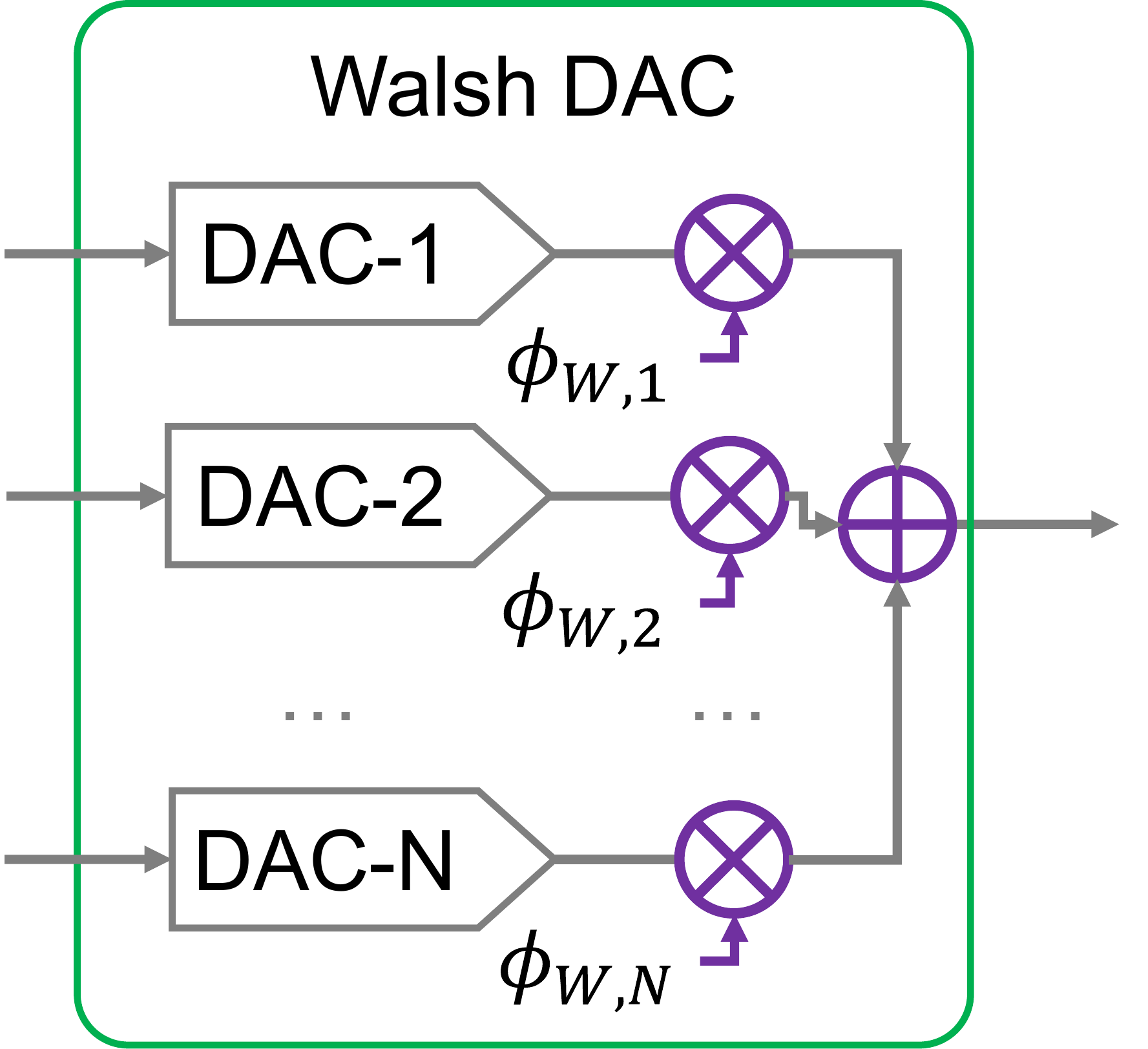}}
\caption{Diagram of Walsh-Hadamard domain ADC (a) and DAC (b), highlighting the differences with respect to time interleaved converters.}
\label{fig:walshconverters}
\end{figure}

Finally, Figure~\ref{fig:walshconverters} shows the adaptation made to interleaved converters to support concurrent WH-domain conversion, which is primarily the mixing in the analog domain with WH basis functions\cite{dehos_d-band_2022, bouassida_concurrent_2016, ferrer_walsh-based_2023, ferrer_experimental_2025} in each sub-converter branch.

\begin{figure*}[tb]
\centering
\includegraphics[width=0.7\linewidth, keepaspectratio]{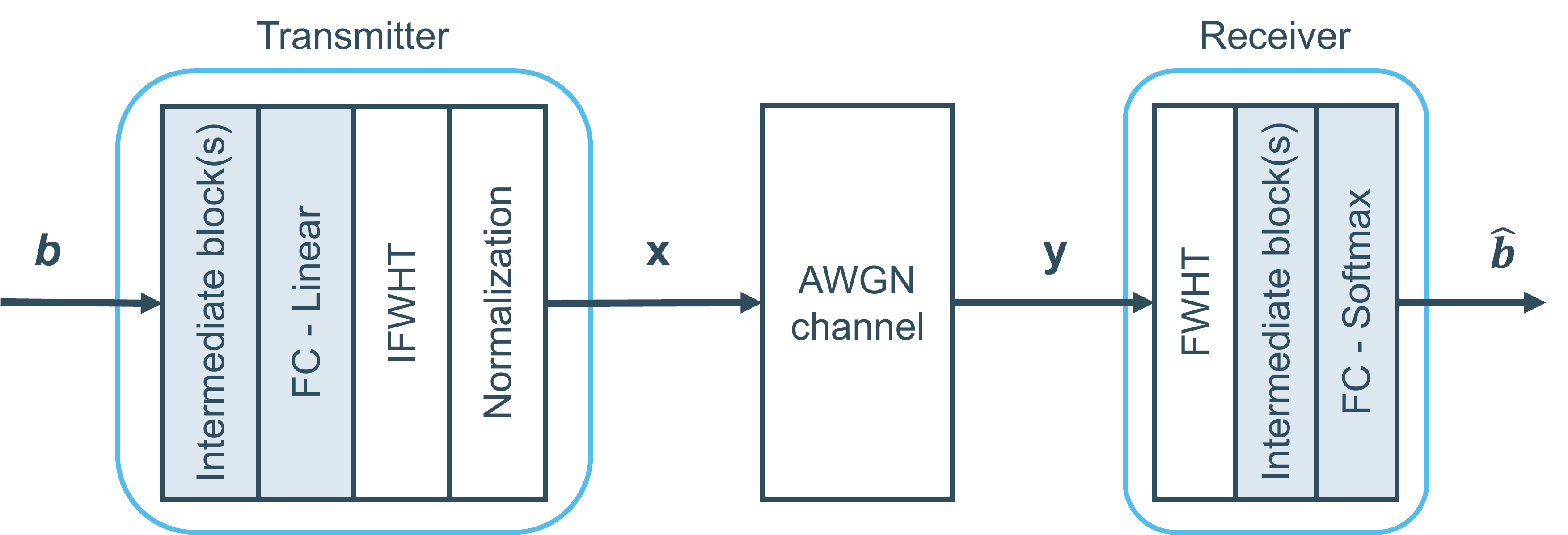}
\caption{Block diagram of the proposed Walsh-Hadamard domain autoencoder, with trainable layers having a shaded background.}
\label{fig:walsh_ae}
\end{figure*}

\textit{\textbf{Walsh-Hadamard transceiver architecture}:} In the proposed architecture, baseband symbol transmission is performed in the orthogonal Walsh-Hadamard domain, rather than in the conventional time or frequency domain. From a communication theory perspective, the orthogonal Walsh-domain transmission between the WH-domain transmitter and the WH-domain receiver is modelled as a discrete-time real-valued baseband communication system, as described in \autoref{subsec:system_model}. The Walsh order $N$, which sets the size of the Walsh transformation, is set to equal the block length: $N=n$, such that $N$ Walsh-domain symbols can be mapped directly on $n$ real-valued baseband channel uses.

At the transmitter, the digital baseband generates $n$ real-valued Walsh-domain symbols per transmission interval of $T$ seconds. These symbols are sent to the WH-domain interleaved DAC~\cite{bouassida_concurrent_2016, ferrer_walsh-based_2023,ferrer_experimental_2025} for digital-to-analog domain conversion. At the same time, this DAC also performs a Walsh-to-time transformation in the analog domain. The resulting sequence of $n$ time-domain samples forms the transmit waveform, which is impaired by the discrete-time AWGN channel~\eqref{eq:awgn_real}. 

At the receiver, the inverse processing is done, i.e., the incoming analog signal is sampled by the WH-domain interleaved ADC~\cite{dehos_d-band_2022}, where each sub-ADC performs projection onto a Walsh-Hadamard basis function and analog-to-digital conversion. The resulting $n$ digital samples represent the received Walsh-domain symbols, which are then processed by the digital baseband receiver.

The benefit of Walsh-Hadamard interleaving is improved converter energy efficiency. At the same time, direct Walsh-Hadamard domain symbol transmission and integration with end-to-end intelligent baseband processing enable joint optimization of modulation, coding, and hardware-aware signal processing, thereby improving overall system energy efficiency.

\section{Autoencoder learned coded modulation}\label{sec:learned_modulation}
This section outlines how a learned coding and modulation system can transparently adapt to the proposed Walsh-Hadamard interleaved wideband transceiver. We expand on the system model of \ref{subsec:system_model} to include trainable layers, describe the detailed architecture, training, and hyperparameter optimisation of the proposed WH-domain autoencoder model.

\subsection{Autoencoder system model}\label{subsec:autoencoder_system_model}
The AI system used to model the wideband transceiver from~\autoref{sec:problem_formulation} is shown in Figure~\ref{fig:walsh_ae}. Neural networks perform the digital baseband operations (joint modulation and coding) in an autoencoder setup, and the functionality of the Walsh-Hadamard domain converter hardware has been abstracted into layers implementing the IFWHT and FWHT~\cite{thys_walsh-domain_2024}. 

The trainable encoder $f_{\theta_T}$ with parameters $\theta_T$ maps the input bits $\boldsymbol{b} \in \mathbb{B}^k$ to Walsh-domain codewords $\boldsymbol{x}_w$. These are transformed to the time domain by means of the IFWHT to form real-valued time domain symbols $\boldsymbol{\tilde{x}} \in \mathbb{R}^{n}$:
\begin{IEEEeqnarray}{rCl}
    \boldsymbol{x}_w & = & f_{\theta_T} (\boldsymbol{b}), \label{eq:encoder}\\
    \boldsymbol{\tilde{x}} & = & \mathcal{W}^{-1}(\boldsymbol{x}_w).
\end{IEEEeqnarray}  

To satisfy the power constraint in Equation~\ref{eq:power_constraint} \cite{oshea_introduction_2017,cammerer_trainable_2020}, the transmitter output $\boldsymbol{\tilde{x}}$ passes through a normalization layer. The constraint is satisfied by normalizing each output to unit power, where the average power is computed during training and reused during inference:
\begin{equation}
    \boldsymbol{x}  =  \frac{\boldsymbol{\tilde{x}}}{\frac{1}{N_B} \sum_{i=1}^{N_B} \left\{ \|\boldsymbol{\tilde{x}}\|^2 \right\}},  \label{eq:normalization}
\end{equation}
where $N_B$ is the batch size during training, equal to the number of bit vectors $\boldsymbol{b}$ simulated in parallel during one training iteration (see also Algorithm~\ref{alg:training}).

The received symbols $\boldsymbol{y}$ are calculated by passing the transmit output $\boldsymbol{x}$ through the AWGN channel model \eqref{eq:awgn_real}. In the ADC, they are transformed back to the Walsh-domain by means of the FWHT, to form the received vector $\boldsymbol{y}_w \in \mathbb{R}^{n}$, which is decoded by the trainable decoder $g_{\theta_R}$ with parameters $\theta_R$:
\begin{IEEEeqnarray}{rCl}
    \boldsymbol{y}_w & = & \mathcal{W}(\boldsymbol{y}),\\
    \hat{\boldsymbol{p}}(\boldsymbol{b}|\boldsymbol{y}_w) & = & g_{\theta_R} (\boldsymbol{y}_w).
\end{IEEEeqnarray}
The vector $ \hat{\boldsymbol{p}}(\boldsymbol{b}|\boldsymbol{y}_w)$ contains the estimated posterior bit probabilities that are used during system training, while \eqref{eq:hard_decision} is used during evaluation or deployment to compute hard decision bit estimates.

Similar to \cite{gruber_dldecoding_2017, kim_communication_2018, jiang_learncodes_2019,jiang_turbo_2019, hesham_coding_2023}, the SNR inside the AWGN layer during training $\gamma_{\text{train}}$ is set a few dB above the Shannon limit for the autoencoder, which can be calculated from the autoencoder information rate $R$ and \eqref{eq:shannon}:
\begin{IEEEeqnarray}{rCl}
    \gamma_{\text{Shannon}}(R) & = & 10 \log_{10} \left(2^{2R} - 1\right),\label{eq:shannon_snr}\\
    \gamma_{\text{train}}(R) & = & \gamma_{\text{Shannon}}(R) + S, \label{eq:snr_train}
\end{IEEEeqnarray}
where $S$ is a hyperparameter called the training SNR offset. Since $\boldsymbol{x}$ is normalized, $\gamma_{\text{train}}$ is controlled via the noise variance \eqref{eq:snr_conversion}.

\subsection{Autoencoder architecture}\label{subsec:architecture}
This section describes in detail the architecture and parameters of the end-to-end autoencoder model shown in Figure~\ref{fig:walsh_ae}. The trainable encoder and decoder are Deep Neural Networks (DNN)\cite{oshea_introduction_2017, rajapaksha_low_2020}, including custom (I)FWHT and normalization layers to model a WH-domain interleaved communication system.

As noted in \cite{rajapaksha_low_2020}, depending on the system parameters $n$ and $k$, the autoencoder \cite{oshea_introduction_2017} will have vastly different complexity in terms of the number of parameters. We include some intermediate hidden layers of fixed size at the transmitter and receiver side, to add representation power to the model, independent of the rate or block length. The parameters $Q$ and $V$ represent the number of neurons and the number of fixed-size hidden layers. The main autoencoder parameters are summarized in Table~\ref{table:hyperparam}, with nominal values as used in the simulations in section~\ref{sec:results}.

\begin{table}[!htbp]
\caption{Main autoencoder hyperparameters. Note that the bold values are the results of hyperparameter selection in section~\ref{subsec:hyperparam}.}
\label{table:hyperparam}
\centering
\begin{tabular}{|c|l|l|}
\hline
\textbf{Symbol} & \textbf{Description} & \textbf{Considered points} \\
\hline
$R$ & information rate & $1/2, 1, 1.5, 2$ \\
\hline
$n$ & channel uses per message & 32 \\
\hline
$k$ & information bits per message & $R \cdot n$ \\
\hline
$Q$ & number of neurons in hidden layers & 100,\textbf{500},1000 \\
\hline
$V$ & number of hidden layers & 1,2,\textbf{4} \\
\hline
$S$ & training SNR offset to Shannon & 0,\textbf{3},6dB \\
\hline
\end{tabular}
\end{table}

The complete architecture overview is provided in Table~\ref{table:architecture}, including the output shape of each layer and the number of neurons in the Fully-Connected (FC) layers. There are $V$ intermediate blocks inside the transmitter and receiver, where each block includes an FC layer with $Q$ neurons and a nonlinear activation layer. The last FC layer in the transmitter has no nonlinear activation, allowing it to place modulation symbols anywhere in the real domain, while the normalization layer handles power normalization. The final output layer of the receiver uses Softmax activation, so the output vectors $\hat{\boldsymbol{p}}$ can be correctly interpreted as independent bit probability estimates. 

\begin{table}[!htbp]
\caption{Walsh-Hadamard domain autoencoder architecture details.}
\label{table:architecture}
\centering
\begin{tabular}{|l|c|c|}
\hline
\textbf{Layer description} & \textbf{Output size} & \textbf{Nb. neurons} \\
\hline
\multicolumn{3}{|c|}{\textbf{Transmitter:}} \\
\hline
Input layer & k & - \\
\hline
$V \times$ Intermediate Block & $Q$ & $Q$ \\
\hline
FC-Linear & $n$ & $n$ \\
\hline
Normalization & $n$ & - \\
\hline
\multicolumn{3}{|c|}{\textbf{Receiver:}} \\
\hline
$V \times$ Intermediate Block & $Q$ & $Q$ \\
\hline
FC-Softmax & $k$ & $k$\\
\hline
\end{tabular}
\end{table}

In Section~\ref{subsec:hyperparam}, the composition of the intermediate blocks is analysed in more detail, investigating the nonlinear activation function, Batch Normalization~\cite{ioffe_batch_2015} (BN), Dropout~\cite{hinton_improving_2012}, and L2 regularization. The parameters $Q$, $V$, and $S$ are also explored to assess their impact on the autoencoder's performance. The analysis of model storage requirement and inference complexity is given in Section~\ref{sec:powermodel}. 

The objective function for autoencoder training is the binary cross-entropy (BCE) between true transmitted bits $\boldsymbol{b}$ and receiver estimated probabilities $\hat{\boldsymbol{p}}$ (adapted from \cite{cammerer_trainable_2020}):
\begin{IEEEeqnarray}{rCl}
\mathcal{L}(\theta_T, \theta_R) & = & \sum_{j=1}^k \mathbb{E}_{\boldsymbol{y}_w, b_j} \{-\log \hat{p}(b_j | \boldsymbol{y}_w) \},\\
& \approx & \tfrac{-1}{N_B} \sum_{i=1}^{N_B}  \sum_{j=1}^k \left\{ b_j \log(\hat{p}(b_j|\boldsymbol{y}_w)) \right. \nonumber\\ 
& &  \left. + (1-b_j) \log(1 - \hat{p}(b_j|\boldsymbol{y}_w)) \right\} \label{eq:loss_batch}
\end{IEEEeqnarray}

This BCE loss function is estimated using batch statistics during batch gradient descent training and is then used to update the model's trainable parameters. For more details on training and evaluation, see Section~\ref{subsec:training}.

\subsection{Baselines for benchmarking}\label{subsec:baselines}
The proposed Walsh-Hadamard transceiver architecture is benchmarked against multiple baselines. For each baseline, different parts of the proposed system are replaced with conventional approaches, and the corresponding adjustments to the power consumption estimation in Section~\ref{sec:powermodel} are made. All baselines are assumed to use time-interleaved conversion instead of the proposed WH-domain interleaving.

\begin{itemize}
    \item {\bf TI-AE}: standard fully connected channel AE~\cite{cammerer_trainable_2020, rajapaksha_low_2020}, without FWHT and IFWHT layers. This model serves as a time-interleaved baseline for the proposed approach, reusing the same architecture (\ref{subsec:architecture}) and training scheme (\ref{subsec:training}).
    \item {\bf CNN-AE}: convolutional neural network AE (CNN-AE)~\cite{hesham_coding_2023}, is the second neural baseline. This is one of the few models in the literature for modulation and coding to provide performance for different information rates. The results were originally simulated for the complex AWGN channel but are presented in this work at half the rate, reflecting the use of the real AWGN channel; see also~\autoref{subsec:system_model}. The CNN architecture offers a different design tradeoff between information rate and complexity compared to the considered fully connected approach. The results are presented for $n=128$ and $P_e=10^{-2}$~\cite{hesham_coding_2023}, while current work focuses on the more challenging case of $n=32, P_e=10^{-3}$.
    \item{\bf Polar}: a conventional modulation and coding scheme, consisting of BPSK modulation and a Polar code \cite{arikan_channel_2009,ercan_error-correction_2017,tavildar_polar_2017,bioglio_design_2021}, with an appended CRC of 6 bits~\cite{tal_list_2015}. To allow for fair comparison, the error correction performance of Polar codes with the same information rate $R=\tfrac{1}{2}$ and block length $n=32$ as the proposed method was simulated using open source code~\cite{tavildar_polar_2017}. The decoding algorithm used is list decoding~\cite{tal_list_2015}, with three list sizes $L$, where a larger $L$ yields better error correction at the cost of increased complexity. The energy efficiency is estimated based on~\cite{ercan_error-correction_2017} (see \autoref{subsec:baseband_power} for details).
    \item{\bf LDPC}: another conventional scheme combining QPSK and LDPC coding. Since LDPC codes are not expected to be used for block length $n=32$, this baseline is only included in the initial SNR performance comparison for $R=\tfrac{1}{2}$ and not in the discussion of energy efficiency. Specifically, 5G NR PDSCH modulation and coding with MCS index 7 and $n=32$, which uses QPSK modulation and LDPC coding, is simulated using Sionna~\cite{sionna2025}. The original spectral efficiency of $1.0273$~\cite[Table 5.1.3.1-1]{3gpp.38.214} was divided by two to account for the real AWGN channel considered in this work. 
\end{itemize}

\subsection{Training and evaluation}\label{subsec:training}
Autoencoder training is performed using the Adam optimizer with an initial learning rate of $0.001$ for up to $E=500$ epochs. Each epoch consists of $T_{enc}=100$ encoder training steps and $T_{dec}=300$ decoder training steps. In every training step, $N_B=5 \times 10^4$ uniformly random bit vectors $\boldsymbol{b}$ and corresponding noise vectors $\boldsymbol{z}$, matching the training SNR, are generated. At the end of each epoch, a validation set of $5 \times 10^4$ random bit vectors is used to monitor training progress. The learning rate is halved if the validation loss does not improve for 20 consecutive epochs, and training is stopped if the learning rate falls below $10^{-10}$.

\begin{table}[!htbp]
\caption{Training related parameters.}
\label{table:trainparam}
\centering
\begin{tabular}{|c|l|l|}
\hline
\textbf{Symbol} & \textbf{Description} & \textbf{Value} \\
\hline
$N_B$ & Batch size & $50000$ \\
\hline
- & Validation set size & $50000$ \\
\hline
$\Delta \sigma$ & Decoder SNR range & $\pm 2~\text{dB}$ \\
\hline
$\eta$ & Learning rate & $10^{-3}$ down to $10^{-10}$ \\
\hline
\end{tabular}
\end{table}

The final training algorithm is summarized in Table~\ref{table:trainparam} and Algorithm~\ref{alg:training}. This is an alternating training algorithm where the encoder and decoder are trained separately in each epoch. The noise variance $\sigma_{\text{train}}$ is set according to \eqref{eq:snr_train} and \eqref{eq:snr_conversion}, depending on the specific training scenario. 
The training SNR is offset from the Shannon SNR by a factor S, since training at the Shannon SNR is quite challenging and may not yield the best performance. What matters more is the threshold SNR at which a trained model meets the BLER requirement of $P_e=10^{-3}$. The training SNR offset $S$ is further explored in \ref{subsec:hyperparam}.

To allow for more robust training of the decoder, the SNR for decoder training is selected uniformly from the range $\left[\sigma_{\text{train}} - \Delta \sigma, \sigma_{\text{train}}+ \Delta \sigma \right]$. The algorithm is similar to the ones used in \cite{aoudia_modelfree_2019,jiang_learncodes_2019,jiang_turbo_2019}. For brevity, we omit the validation stage and the learning rate update mechanism. 

\begin{algorithm}[htbp] 
\caption{Alternating Training Algorithm} \label{alg:training}
\begin{algorithmic}%[1] 
    \Require Batch Size $N_B$, Encoder Iterations $T_{enc}$, Decoder Iterations $T_{dec}$, Number of Epochs $E$,  Decoder SNR range $\Delta\sigma$, Bits per message $k$, Block length $n$, Learning Rate $\eta$
    \While{$e \leq E$} %AND $ \eta > 10^{-10}$}
        \For{$t \leq T_{enc}$}
            \For{$i \leq N_B$}
                \State Generate random input $\boldsymbol{b} \sim \mathcal{U}(\mathbb{B}^k)$
                \State Generate random noise $\boldsymbol{z} \sim \mathcal{N}(0, \sigma_{\text{train}}^2 \boldsymbol{I}_n)$
                \State Transmission: $\boldsymbol{\hat{b}} = g_{\theta_R}(f_{\theta_T}(\boldsymbol{b})+\boldsymbol{z})$
            \EndFor
            \State Update encoder: $\theta_T = \theta_T - \eta \nabla_{\theta_T} \mathcal{L}$ 
        \EndFor
        \For{$t \leq T_{dec}$}
            \For{$i \leq N_B$}
                \State Generate random input $\boldsymbol{b} \sim \mathcal{U}(\mathbb{B}^k)$
                \State Generate SNR value $\sigma_{\text{dec}} \sim \mathcal{U}(\sigma_{\text{train} } \pm \Delta\sigma)$
                \State Generate random noise $\boldsymbol{z} \sim \mathcal{N}(0, \sigma_{\text{dec}}^2 \boldsymbol{I}_n)$
                \State Transmission: $\boldsymbol{\hat{b}} = g_{\theta_R}(f_{\theta_T}(\boldsymbol{b})+\boldsymbol{z})$
            \EndFor
            \State Update decoder: $\theta_R = \theta_R - \eta \nabla_{\theta_R} \mathcal{L}$ 
        \EndFor
%        \State Generate random validation data $\boldsymbol{b} \sim \mathcal{U}(\mathbb{B}^k)$
%        \State Compute validation loss $\mathcal{L}_{val}(\theta_T, \theta_R)$
%        \State If $\mathcal{L}_{val}$ did not improve for 20 epochs: $\eta = \eta / 2$
    \EndWhile
\end{algorithmic}
\end{algorithm}

%% AI training details
In Section~\ref{sec:results}, we train several WH-AE and TI-AE models with different coding rates using algorithm~\ref{alg:training}. Training for one autoencoder is performed on a single Nvidia 2080 Ti GPU and is limited to 2 days or 500 epochs, whichever comes first. After training, the BLER for each trained model is estimated via Monte Carlo simulation over a large number of random bit vectors $\boldsymbol{b}$ and across a range of evaluation channel SNRs $\gamma$. From these BLER-SNR curves, we extract the threshold SNR for a target BLER of $10^{-3}$ according to \eqref{eq:threshold_snr}, which is the primary error correction performance indicator.

\begin{figure}[htb]
\centering
\includegraphics[width=\linewidth, keepaspectratio]{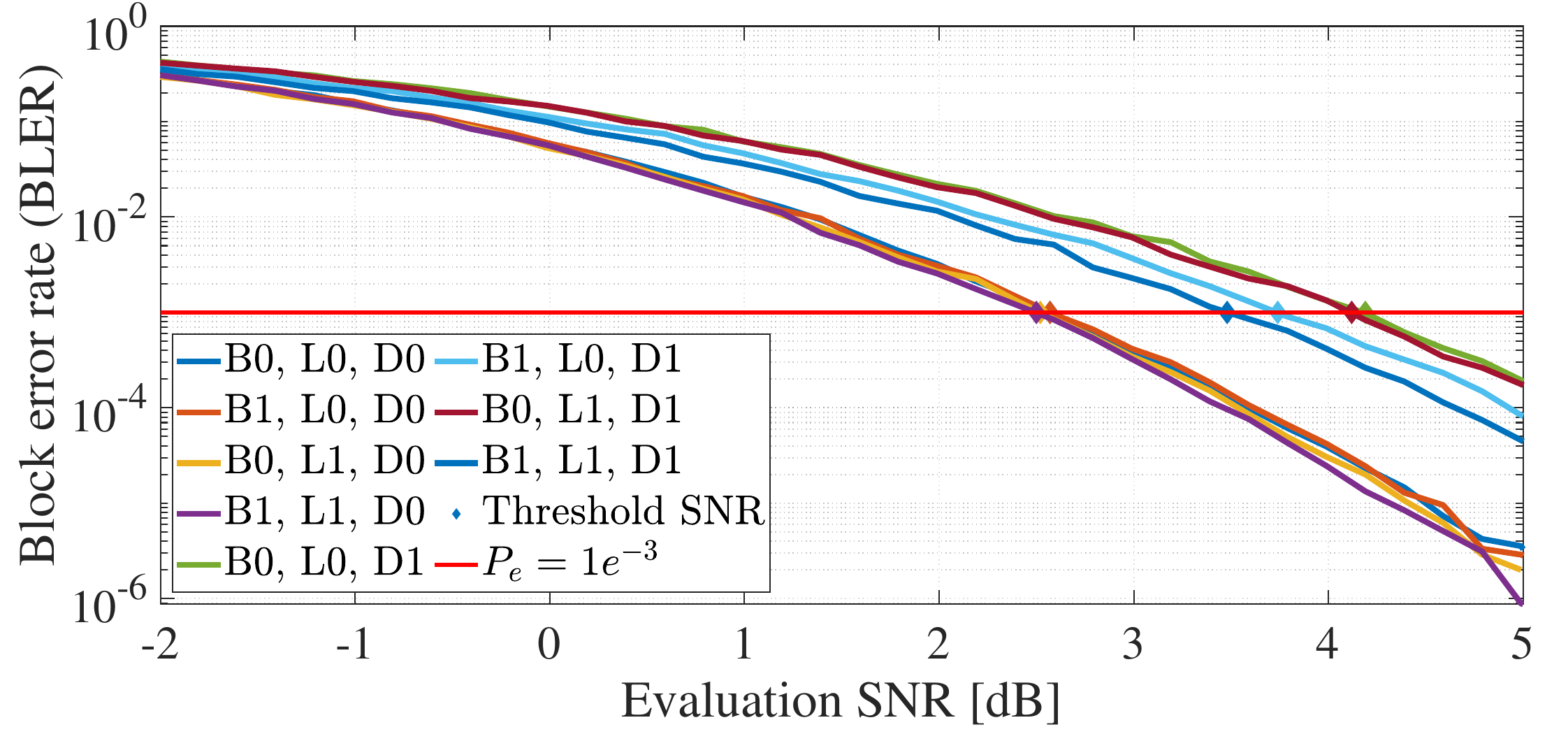}
\caption{Example BLER curves for autoencoders with $N=32$, $k=16$. The horizontal line indicates the BLER threshold $P_e=0.1\%$. The data is from Section~\ref{subsec:hyperparam}, with L2 regularization enabled. The threshold SNR for each model is indicated with a diamond on the figure.}
\label{fig:bler_example}
\vspace{-2mm} % to reduce white space after figure
\end{figure}

Figure~\ref{fig:bler_example} shows an example of model evaluation using a Monte Carlo simulation. The results for several models (from \ref{subsec:hyperparam}) are shown in this figure, allowing their error correction performance to be compared. From these BLER curves, the threshold SNR to reach $P_e=0.1\%$ is extracted, where a lower threshold SNR indicates a higher performing coding.

\subsection{Hyperparameter selection} \label{subsec:hyperparam}
To determine the best architecture for the proposed WH-AE, a hyperparameter grid search is performed with a fixed scenario of $n=32, k=16$ ($R=\tfrac{1}{2}$). The final architecture will be used in Section~\ref{sec:results} to train our proposed AE models in the Walsh-Hadamard domain with varying code rates. The hyperparameter search consists of two stages: first, optimising the architecture of the intermediate hidden layers; and second, optimising the number of neurons $Q$, the number of layers $V$, and the training SNR offset $S$. 

The best configuration at each stage is selected based on the threshold SNR \eqref{eq:threshold_snr}, with a lower threshold SNR being better. It should be noted that for the specific $R=\tfrac{1}{2}$ case, the Shannon SNR is $0$ dB and thus the training SNR is equal to the training SNR offset: $\gamma_{\text{train}}=0+S$. 

\textit{\textbf{Optimising hyperparameters for threshold SNR}:} The hidden layer architecture is optimized over 4 binary decisions: whether to include batch normalization~\cite{ioffe_batch_2015}, choosing between ReLU and LeakyReLU activation, Dropout~\cite{hinton_improving_2012}, and L2 weight regularization. L2 regularization is an additional loss term that penalizes large weights and helps improve convergence during training. When L2 regularization is enabled, it is applied to all Fully Connected layers in the model.

The first stage hyperparameter search is performed with the parameters $Q,V,S$ fixed to $500$, $4$ and $3$ dB, respectively. Fig.~\ref{fig:bler_example} shows the BLER - SNR curves for models trained with L2 regularization, where $B,L \text{ and } D$ represent batch normalization, LeakyReLU, and Dropout. The threshold SNR to reach $P_e=0.1\%$ is extracted from these curves. One can already discern from Figure~\ref{fig:bler_example} that models with Dropout ($D1$ on the graph) are performing poorly.

\begin{figure}[!htb]
\centering
\includegraphics[width=\linewidth, keepaspectratio]{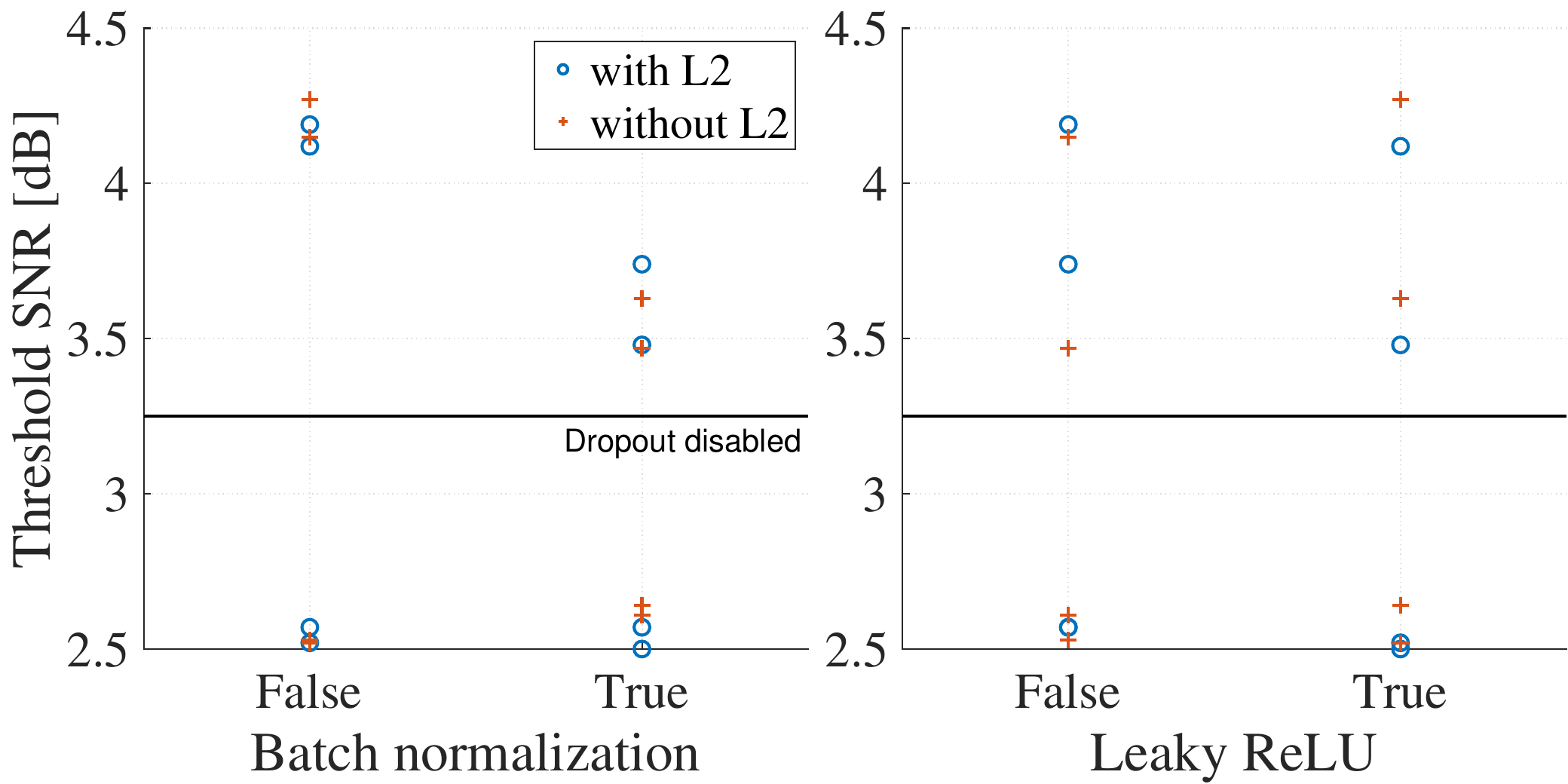}
\caption{Results of the first stage hyperparameter search.}
\label{fig:arch_search}
\vspace{-2mm} % to reduce white space after figure
\end{figure}

The main finding of Fig.~\ref{fig:bler_example} is that enabling Dropout has a sizeable negative impact on the error correction performance. Figure~\ref{fig:arch_search} shows an alternative visualization of the first hyperparameter search, focusing only on the threshold SNR \eqref{eq:threshold_snr}.

The figure shows negligible variance in the results across the batch normalization, activation, and L2 regularization parameters. Similar to Fig.~\ref{fig:bler_example}, there is a significant reduction in threshold SNR for the architectures without Dropout. In fact, all results with Dropout enabled achieve a threshold SNR of around $3.5\,dB$ or above. The minimal threshold SNR at this stage is achieved without Dropout, but with batch normalization, LeakyReLU activation, and L2 enabled.

\begin{figure}[htb]
\centering
\includegraphics[width=\linewidth, keepaspectratio]{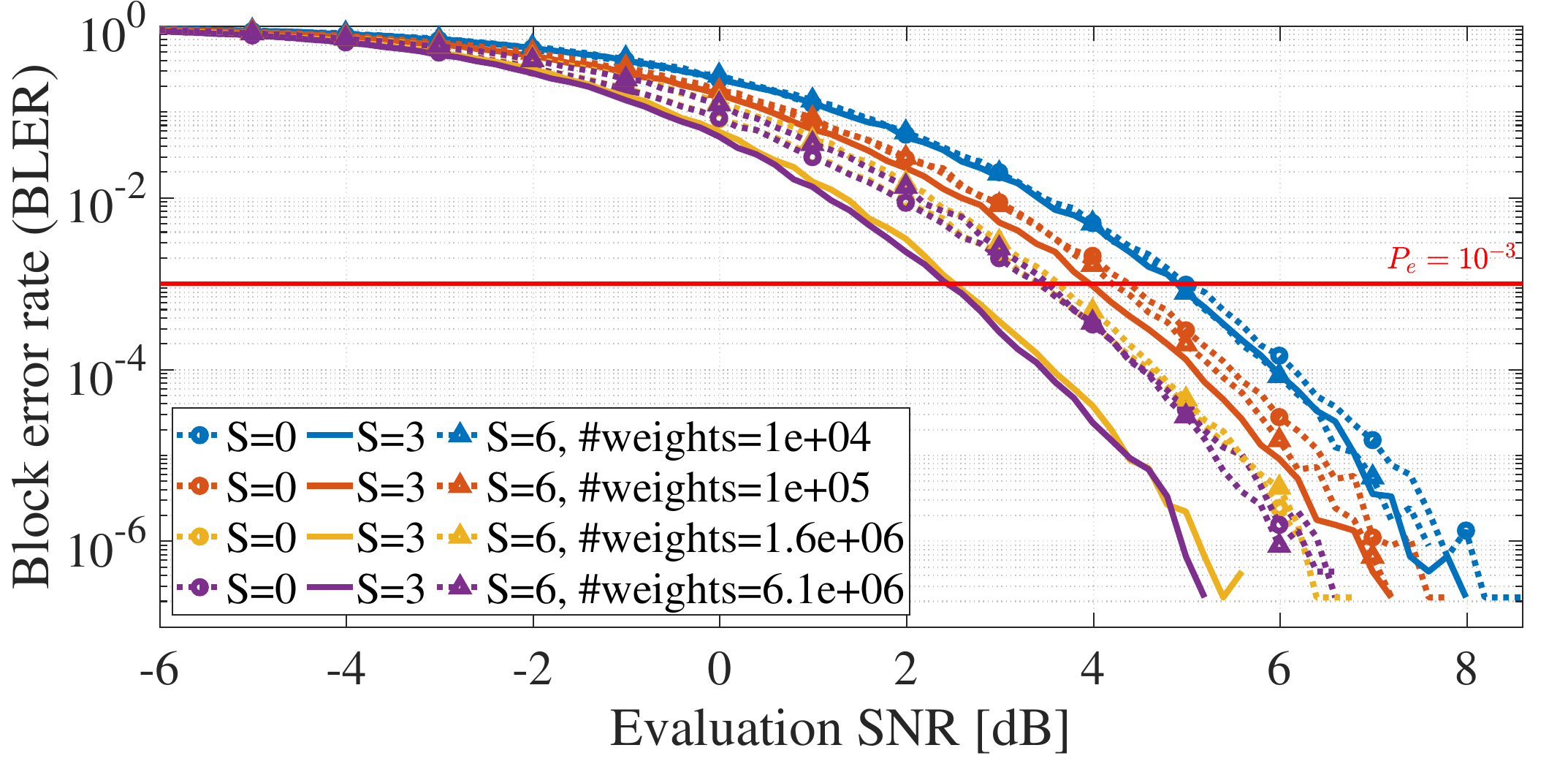}
\caption{BLER curves for autoencoders with $N=32$, $R=\tfrac{1}{2}$. The horizontal line indicates the BLER threshold $P_e=0.1\%$. All rows in the legend are models with the same number of weights (determined by parameters $Q,V$), with the differentiating factor being the training SNR offset $S$ in dB.}
\label{fig:bler_example_qvs}
\vspace{-2mm} % to reduce white space after figure
\end{figure}

\begin{figure}[!htb]
\centering
\includegraphics[width=\linewidth, keepaspectratio]{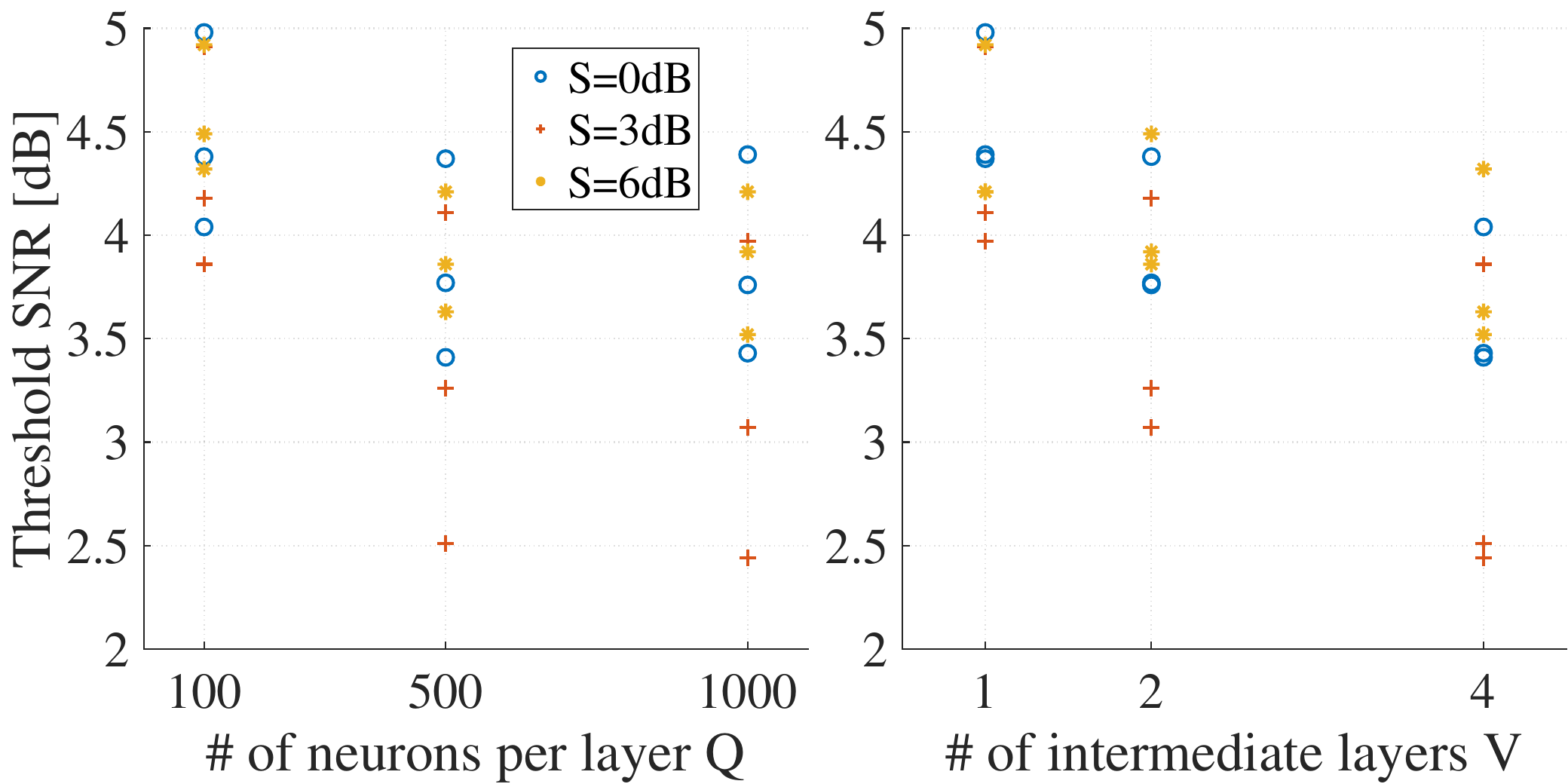}
\caption{Results of the second stage hyperparameter search. The two subplots are separate views on the same dataset, e.g., looking at $Q=500$, there are 9 distinct results corresponding to 3 choices for $S$ and 3 choices for $V$.}
\label{fig:qvs_search_results}
\vspace{-2mm} % to reduce white space after figure
\end{figure}

The second stage hyperparameter search is performed with the best architecture from the first stage, while the number of hidden layers $V$, the number of neurons per layer $Q$, and the training SNR offset $S$ are varied over the ranges in Table~\ref{table:hyperparam}. We perform a small grid search over the parameters $Q,V,S$ and show a selection of the resulting BLER curves in Figure~\ref{fig:bler_example_qvs} (a small subset is shown for clarity). All threshold SNR results are further reported in Fig.~\ref{fig:qvs_search_results}.

The figure shows the impact of $S$, the training SNR offset from the Shannon SNR, on the BLER performance of trained autoencoder models. Especially for high complexity models with a large number of weights, the optimal choice for $S$ appears to be $3\,dB$. Consider, for example, the model with $6.1e+06$ weights (corresponding to $Q=1000$ neurons per layer and $V=4$ layers): using $S=3\,dB$ improves the threshold SNR by a whole decibel compared to using $S=0$ or $6\,dB$.

We can now analyze the behavior of the threshold SNR with the hyperparameters $Q,V \text{ and } S$. The results are shown in Figure~\ref{fig:qvs_search_results}, which shows that error correction performance scales with increasing model complexity. The best threshold SNR for any combination of $Q,V$ is achieved using a training SNR offset $S = 3\,dB$.

\textit{\textbf{Impact on training convergence}:} To further assess the impact of hyperparameter selection on error correction performance, it is instructive to look deeper at the effects on training behavior. Validation accuracy is typically used in classification tasks as an indication of training convergence. Therefore, we analyse here the final validation accuracy after the last training epoch. The validation accuracy is measured by simulating the trained model on a randomly generated batch of binary data at the training SNR \eqref{eq:snr_train}, and thus reflects the model's accuracy at that SNR. Hence, we compare the AI model's convergence on the classification task used for training with its error correction performance, the actual performance metric of interest at inference.

\begin{figure}[htb]
\centering
\includegraphics[width=\linewidth, keepaspectratio]{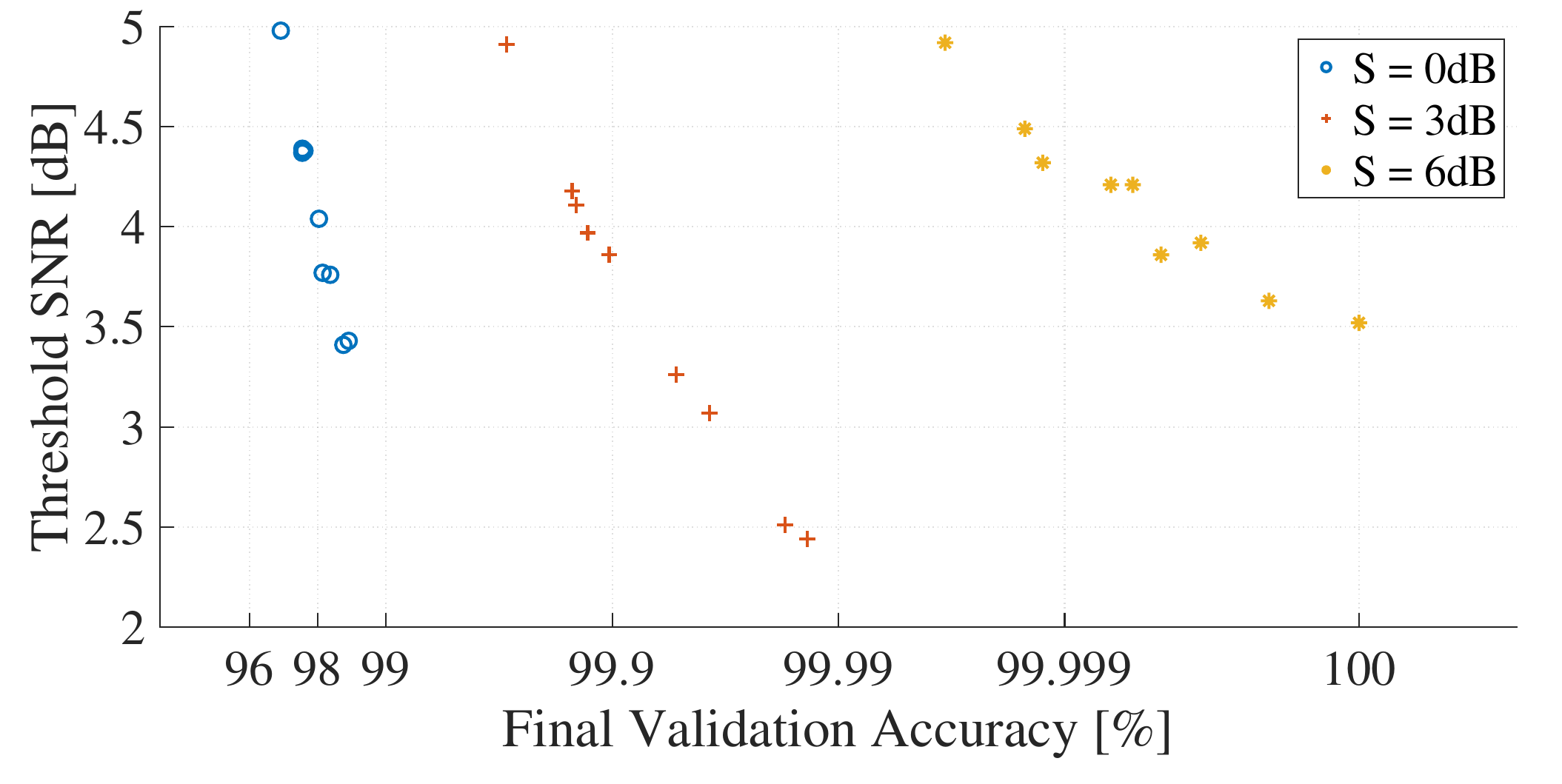}
\caption{Impact of training SNR offset $S$ on error correction performance and training convergence.}
\label{fig:layer_search_SNR_vs_ACC} 
\vspace{-2mm} % to reduce white space after figure
\end{figure}

The comparison in Figure~\ref{fig:layer_search_SNR_vs_ACC} is made for the second stage hyperparameter optimisation results, varying the parameters $Q,V \text{ and } S$. 

The results in Fig.~\ref{fig:layer_search_SNR_vs_ACC} show clear grouping according to the training SNR offset $S$. The error correction performance within a group improves proportionally as the model complexity increases with the parameters $Q$ or $V$. 

Furthermore, models trained with $S=6\,dB$ achieve up to $100\%$ validation accuracy, suggesting they converge during training, but they do not achieve the best overall threshold SNR. This effect is due to the rarity of error events when training at large SNR, leading to batches of data bits without a single error and a biased validation accuracy.

The data in this section indicate that the training SNR $S$ has a significant impact on convergence and threshold SNR performance, with an optimum at $ S=3\,dB$ above the Shannon limit. Indeed, the absolute minimum threshold SNR can be achieved with $S=3\,dB$ using the most complex model, which contains 4 intermediate layers and 1000 neurons per layer. 

Up to now, we have analysed the impact of hyperparameters on error correction performance, measured as threshold SNR, and on training convergence, using the validation accuracy metric. A careful analysis of the AI-based coded modulation system's complexity and power consumption is now needed to determine Pareto-optimal hyperparameters, which is the focus of the following section.

\section{System-level power and energy model} \label{sec:powermodel}
We propose a system-level power model to compare the energy efficiency of joint modulation and coding designs, including the proposed Walsh-Hadamard transceiver architecture. The model consists of two parts: the power consumption of the baseband and the power consumption of the converter hardware.
\begin{equation}
    P_{\text{sys}} = P_{\text{BB}} + P_{\text{ADC}}+P_{\text{DAC}}.\label{eq:system_power}
\end{equation}
Before defining the system-level energy efficiency, the following sections detail each independent part of the power model.

\subsection{Converter power consumption}
The converter power consumption varies depending on the system considered: while the proposed solution is based on WH-interleaved converters, other baselines use time-interleaved (TI) converters. For each type of converter, a representative power consumption is listed in Table~\ref{table:converterpower}:

\begin{table}[!htbp]
\caption{Converter power consumption}
\label{table:converterpower}
\centering
\begin{tabular}{|c|l|l|l|}
\hline
\textbf{Reference} & \textbf{Type} & \textbf{Specification} & $P$\,[mW] \\
\hline
\cite{ferrer_walsh-based_2023} & Walsh DAC & 8b @ 5\,GS/s & $44$ \\
\hline
\cite{radulov_28-nm_2015} & TI DAC & 6b @ 7\,GS/s & $145$ \\
\hline
\cite{dehos_d-band_2022} & Walsh ADC & 6b @ 5\,GS/s & $46$ \\
\hline
\cite{ramkaj_33_2019} & TI ADC & 12b @ 5\,GS/s  & $158.6$ \\
\hline
\end{tabular}
\end{table}

Since~\cite{ferrer_walsh-based_2023,dehos_d-band_2022} are the only published WH domain converters, the TI converter baselines are selected with comparable symbol rate and resolution (marked 'specification' in Table~\ref{table:converterpower}).

\subsection{Baseband power consumption estimation}\label{subsec:baseband_power}
The baseband power consumption is estimated based on the following formula:
\begin{equation}
    P_{\text{BB}}=E_{\text{BB}} f_{\text{BB}},
\end{equation}
where $E_{\text{BB}}$ is the baseband energy expenditure, and the baseband sampling rate for the considered wideband communication scenario is set as follows: 
\begin{equation}
f_{\text{BB}} = \frac{f_s}{n} = \frac{5\,\text{GSPS}}{32} = 156.25\,\text{MHz}, 
\end{equation}
where an interleaving order $n=32$ is selected for all results. For numerical details on the power consumption of the baselines, see the explanation below.

\textit{\textbf{AI inference power consumption}:} To model AI inference power consumption, we perform a detailed analysis of all computations in the AI baseband, which we divide by an empirical efficiency factor (in Operations-per-Joule) for a representative hardware platform to calculate the overall energy expenditure for each AI model.

Specifically, arithmetic performed in each neural network layer is counted as OPerations (OPs), and the total sum of OPs $C_{\text{AI}}$ for one model is then used to estimate that model's computational complexity. To estimate energy expenditure from computational complexity, we define the following:
\begin{IEEEeqnarray}{lCl}
    E_{\text{BB,AI}} & = & \frac{C_{\text{AI}}}{\eta}, \label{eq:ai_power}\\
    C_{\text{AI}} & = & \sum\limits_{j=1}^{\#\text{layers}} C_{j}(I_j,O_j). \label{eq:complexity}
\end{IEEEeqnarray}
Here $C_{j}(I_j,O_j)$ is the complexity of the j-th layer with respective input/output size $I_j$, $O_j$. Furthermore, $\eta$ is the AI hardware accelerator's compute efficiency in OP per Joule (equivalently, in OP per second (OPS) per Watt). In the current work, we select a highly optimized 22-nm FD-SOI accelerator for low-precision neural network inference, with a hardware compute efficiency of 800~TeraOPS-per-Watt~\cite{kneip_impact_2023}. This choice is motivated by the high complexity of current channel AE systems, which require aggressive quantization, weight pruning, and high-performance hardware to achieve real-time operation~\cite{wiesmayr_design_2024}.

The complexity for each layer type is listed in Table~\ref{table:formulas} below, where $I$ and $O$ represent the input and output feature vector sizes. The complexity of (I)FWHT layers is not counted since these are only present during training; at inference time, these layers are functionally replaced by the Walsh converter hardware, whose power consumption is already accounted for in $P_{\text{sys}}$. The batch normalization (BN) and normalization (NORM) layers operate element-wise, so $I=O$ in those layers. The exact operation count for each layer in the architecture is given in Table~\ref{table:architecture}.

\begin{table}[!htbp]
\caption{Complexity \& Memory footprint of neural network layers}
\label{table:formulas}
\centering
\begin{tabular}{|l|c|c|}
\hline
\textbf{Layer type} & \textbf{Complexity} $C(I,O)$ & \textbf{Memory} $M(I,O)$ \\
\hline
Fully Connected & $2IO$ & $(I+1)O$ \\
\hline
Batch normalization & $4I$ & $4I$ \\
\hline
Normalization & $I$ & $I$ \\
\hline
Convolution & $(2 C F+1)I O/S - 1$ & $(CF+1)O$\\
\hline
\end{tabular}
\end{table}

To estimate the model's memory footprint, Table~\ref{table:formulas} also lists the number of parameters $M$ per layer. The BN layer has 4 state parameters per output channel: the learnable scale and offset, and the moving average mean and variance \cite{hinton_improving_2012}. The NORM layer has 1 state parameter per output channel, which is the normalization factor \eqref{eq:normalization}.

To summarize the analysis of memory footprint and complexity, Table~\ref{table:complexity} provides the calculated number of parameters $M$ and operations $C$ for all layers in the proposed architecture, as shown in Table~\ref{table:architecture}. The results are parameterized by the hyperparameters $Q,V$, in the block length $n$ and the number of information bits $k$.

\begin{table}[!htbp]
\caption{Memory \& Complexity of proposed autoencoder architecture}
\label{table:complexity}
\centering
\begin{tabular}{|l|c|c|}
\hline
\textbf{Layer description} & \textbf{Parameters} $M$ & \textbf{Operations} $C$ \\
\hline
\multicolumn{3}{|c|}{\textbf{Transmitter:}} \\
\hline
$V \times$ &  $Q(k+1) + $ & $2kQ + $\\
Intermediate Block & $(V-1)(Q^2+Q)$ & $(V-1)2Q^2$ \\
\hline
FC-Linear & $n (Q+1)$ & $2 Q n$ \\
\hline
Normalization & n & n \\
\hline
\multicolumn{3}{|c|}{\textbf{Receiver:}} \\
\hline
$V \times$ &  $Q(k+1) + $ & $2kQ + $\\
Intermediate Block & $(V-1)(Q^2+Q)$ & $(V-1)2Q^2$ \\
\hline
FC-Softmax & $k(Q+1)$ & $ 2 Q k$ \\
\hline
\end{tabular}
\end{table}

\textit{\textbf{Baseline power consumption}:} The power consumption of each baseline considered in this work is also analyzed. All baselines are assumed to use a time-interleaving domain conversion architecture, so the converter power consumption follows those marked as 'TI' in \autoref{table:converterpower}. The TI-AE baseline uses the same AI architecture as the proposed system and thus the same baseband power model, with the only difference being the domain converter's power consumption.

The CNN-AE baseline power consumption is estimated by including the complexity of the convolutional layers, for which we calculate the complexity and memory impact in Table~\ref{table:formulas}. In these equations, $I$ represents the CNN input size, $O$ the number of output channels, $C$ the input channels, $F$ the filter size, and $S$ the stride. The CNN-AE layer architecture is described in detail in~\cite{hesham_coding_2023}.

For the conventional modulation and coding baseline, we assume a TI transceiver with BPSK modulation and a Polar code. Since Polar decoding is expected to dominate energy consumption, we use the energy efficiency figures as reported in \cite{ercan_error-correction_2017} for 'Fast SSCL' Polar decoders as a value for $E_{\text{BB, Polar}}$. Since there are no publicly reported figures for block length $n=32$, we scale the results in \cite{ercan_error-correction_2017} for $n=256$, by assuming a linear scaling of decoding complexity with block length. The 'Fast SSCL' algorithm is a specific variant of Polar list decoding, with performance identical to that of the decoding algorithm used in the baseline simulations~\cite{ercan_error-correction_2017,tavildar_polar_2017,tal_list_2015}.

\subsection{Energy efficiency model}
Finally, the system-level energy efficiency is defined as the information bit throughput over the system power consumption~\eqref{eq:system_power}. It is measured in bits per second per Watt or bits per Joule:
\begin{align}
    EE_{\text{sys}} & = \frac{k \cdot f_{\text{BB}}}{P_{\text{sys}}}=\frac{k}{E_{\text{sys}}}. \label{eq:efficiency}
\end{align}

\begin{figure}[!htbp]
\centering
\includegraphics[width=\linewidth, keepaspectratio]{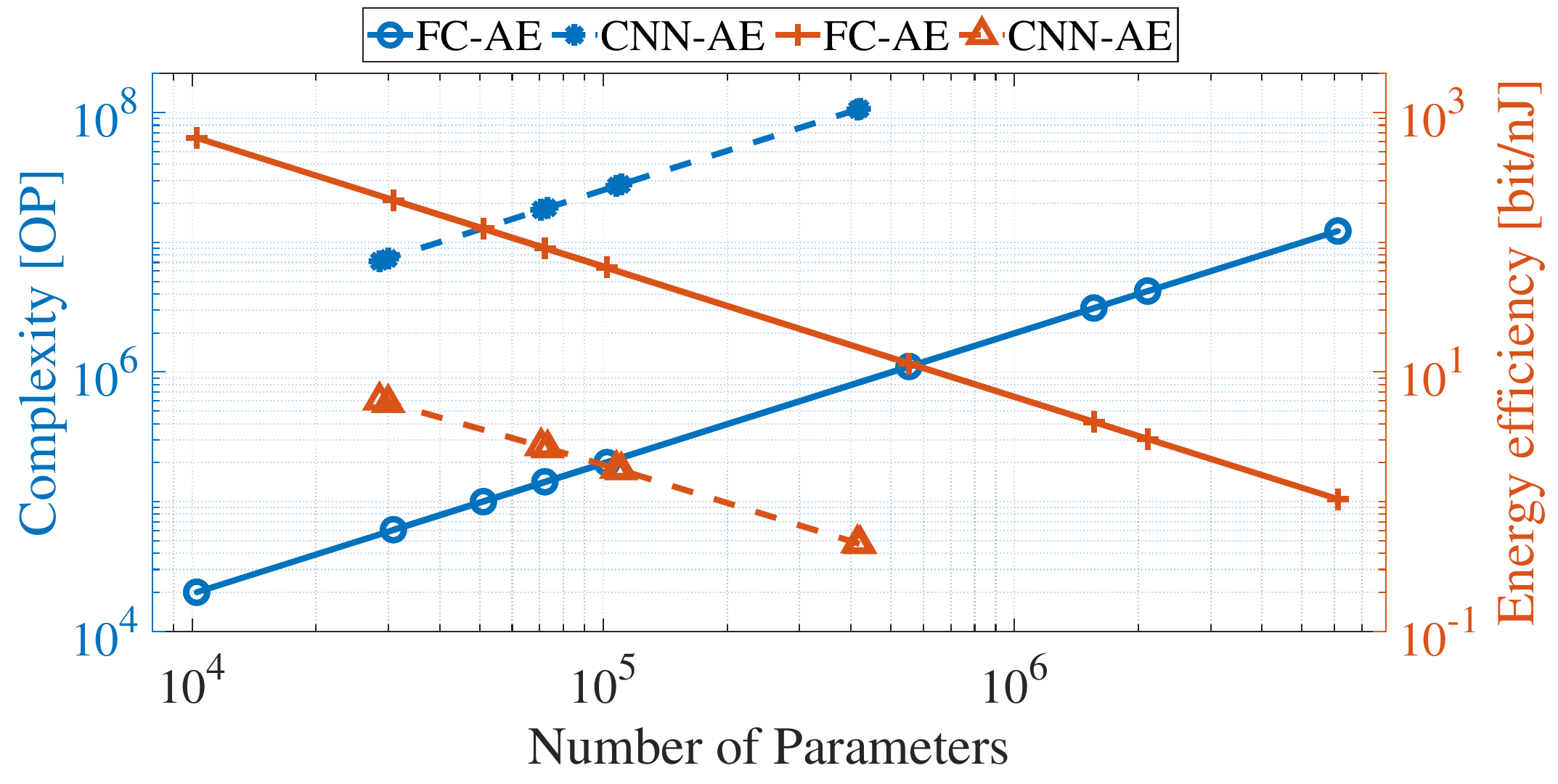}
\caption{Computational complexity and energy efficiency as a function of model size ($n=32$, $R=1/2$). To focus solely on the AI architecture impact, energy efficiency excludes converter energy.}
\label{fig:power_model}
\vspace{-2mm} % to reduce white space after figure
\end{figure}

To illustrate the impact of AI architecture on energy efficiency, Fig.~\ref{fig:power_model} shows the model complexity \eqref{eq:complexity} and energy efficiency \eqref{eq:ai_power} as a function of model size. It compares the two investigated AI architectures: the proposed AE with FC layers (FC-AE) and the baseline CNN-AE. The model size is swept by varying the architecture-specific hyperparameters ($Q$ and $V$ for FC-AE) while keeping the scenario fixed. The FC-AE model is selected with $n=32, k=16$ and BN enabled, while for the CNN-AE architecture, $n=128, k=64$ was used, as in~\cite{hesham_coding_2023}.

The graph shows complexity grows linearly with model size while energy efficiency is inversely proportional to model size, given the fixed number of information bits $k$. While the CNN-AE boasts a relatively low parameter count (especially given the larger block length), its computational complexity and energy efficiency are worse than those of the FC-AE. A complete comparison will also take into account error correction performance, so the full tradeoff is detailed in section~\ref{subsec:full_tradeoff}.

\subsection{Pareto optimal hyperparameter selection}\label{subsec:pareto_hyperparam}
\begin{figure}[htb]
\centering
\includegraphics[width=\linewidth, keepaspectratio]{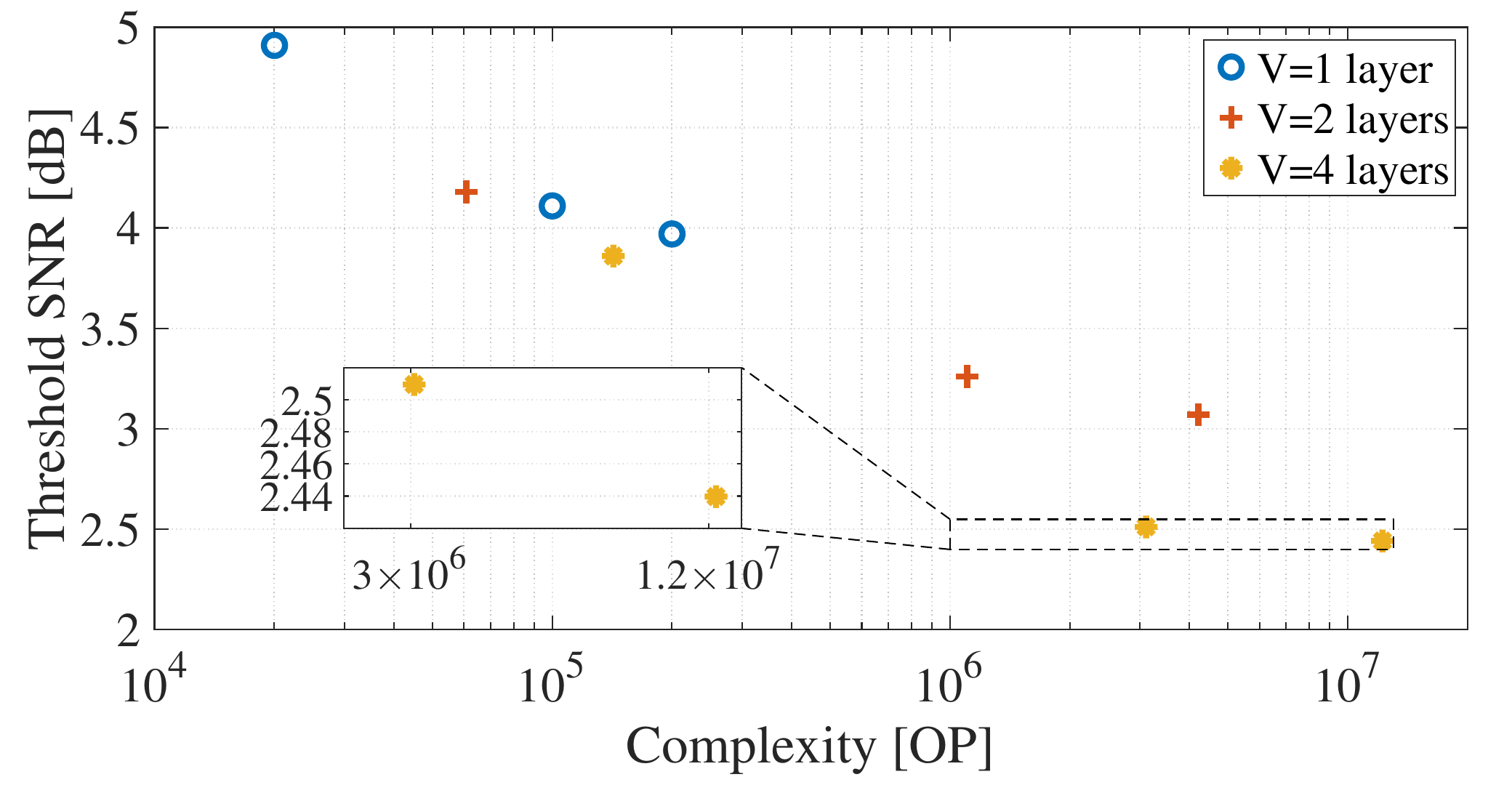}
\caption{Trade-off between error correction performance and implementation complexity for $n=32, R=\frac{1}{2} \text{ and } P_e=0.1\%$. For each value for the number of layers $V$, the complexity grows proportional to the number of neurons $Q=\{100, 500,1000\}$.}
\label{fig:snr_complexity}
\vspace{-2mm} % to reduce white space after figure
\end{figure}

We are now in a position to finalize the hyperparameter selection from \ref{subsec:hyperparam}. We analyse the complexity-performance tradeoff for the rate $\tfrac{1}{2}$ scenario in Fig.~\ref{fig:snr_complexity}. As found in the analysis in \ref{subsec:hyperparam}, we use batch normalization, L2 regularization, and LeakyReLU activations, with the training SNR offset $S$ fixed at $ 3\,dB$. The complexity in OPs, calculated using \eqref{eq:complexity}, scales with increasing number of neurons $Q$ and number of layers $V$. The datapoints are grouped by the number of layers $V$.

Figure~\ref{fig:snr_complexity} shows that the best model for error correction is the most complex one, but also that there is a drastic reduction in complexity available by selecting the second-best architecture. This corresponds to choosing between a 4-layer autoencoder with 500 or 1000 neurons per layer, which seems to have diminishing returns in terms of error correction performance. Indeed, an increase of only $ 0.07$ dB in threshold SNR allows an order-of-magnitude reduction in model complexity, from 12 to 3 GigaOP per inference. 

To conclude the hyperparameter optimisation, the final architecture used in the following simulations is a 4-layer autoencoder with $Q=500$ neurons per layer, batch normalization, LeakyReLU activation, L2 weight regularization, and a training SNR set to $ 3\,dB$ above the Shannon limit for the considered information rate.

\section{Results}\label{sec:results}
After outlining our proposed method in Section~\ref{sec:learned_modulation}, we analyse the results and compare them with the baselines from Section~\ref{subsec:baselines}. With hyperparameter optimisation taken care of in~\ref{subsec:hyperparam}, the proposed WH-AE architecture is compared to time-interleaved baselines in terms of threshold SNR, achievable information rate, total system power consumption, and energy efficiency. For reference, the system parameters are listed in Table~\ref{table:hyperparam}, and the model training details are presented in Table~\ref{table:trainparam}. As a first step, we analyse the reliability and energy efficiency of the proposed WH-AE coded modulation system for a single information rate. Finally, we analyse the scaling of energy efficiency with information rate for both the baseline neural coded modulation systems and the proposed WH-AE.

\subsection{Performance in single rate scenario} \label{subsec:results_reliability}
\begin{figure}[!htbp]
\includegraphics[width=\linewidth, keepaspectratio]{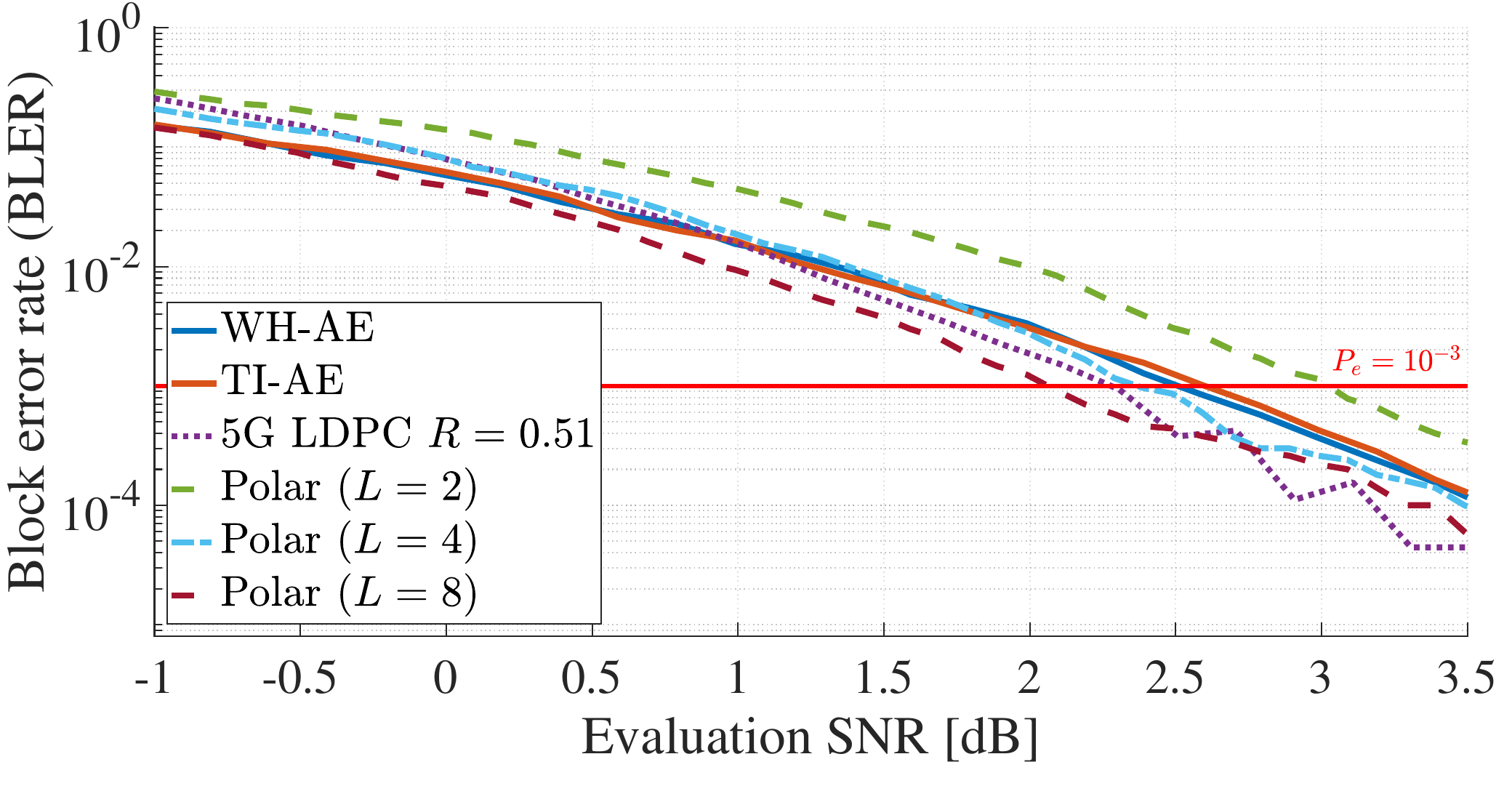}
\caption{Reliability of the proposed system and selected baselines, with block length $n=32$ and rate $R=\tfrac{1}{2}$. The threshold SNR is extracted by finding the intersection of the BLER-SNR curves with the horizontal line indicating the BLER threshold of $P_e=0.1\%$.}
\label{fig:blercurves}
\vspace{-2mm} % to reduce white space after figure
\end{figure}

Focusing first on a single information rate, Fig.~\ref{fig:blercurves} shows the results for models with $n=32$ and $R=\frac{1}{2}$, comparing WH-domain autoencoders, with optimal hyperparameters, to the baselines in terms of BLER. The Polar baseline results for list sizes $L=2,4,8$ confirm that larger $L$ leads to better error-correction performance. The CNN-AE baseline is not included since the authors of \cite{hesham_coding_2023} do not report the BLER performance over an SNR range. The results reinforce the fact that finite block length coding systems operate far from the Shannon capacity limit, which is $0\,dB$ in this case (see also Subsection~\ref{subsec:system_model}). The proposed WH-AE autoencoder is competitive with all baselines in this scenario, achieving reliability similar to that of the TI-AE and Polar with $L=4$, while the best reliability is achieved by Polar with $L=8$.

We now analyze the system power consumption of the proposed Walsh-Hadamard autoencoder transceiver architecture, following the methods outlined in Section~\ref{sec:powermodel} (Eqn.~\ref{eq:system_power}). The wideband converter power consumption is listed in Table~\ref{table:converterpower}, with the baselines using time-interleaved conversion and the proposed method using Walsh-Hadamard domain conversion. 

\begin{figure}[htb]
\centering
\includegraphics[width=\linewidth, keepaspectratio]{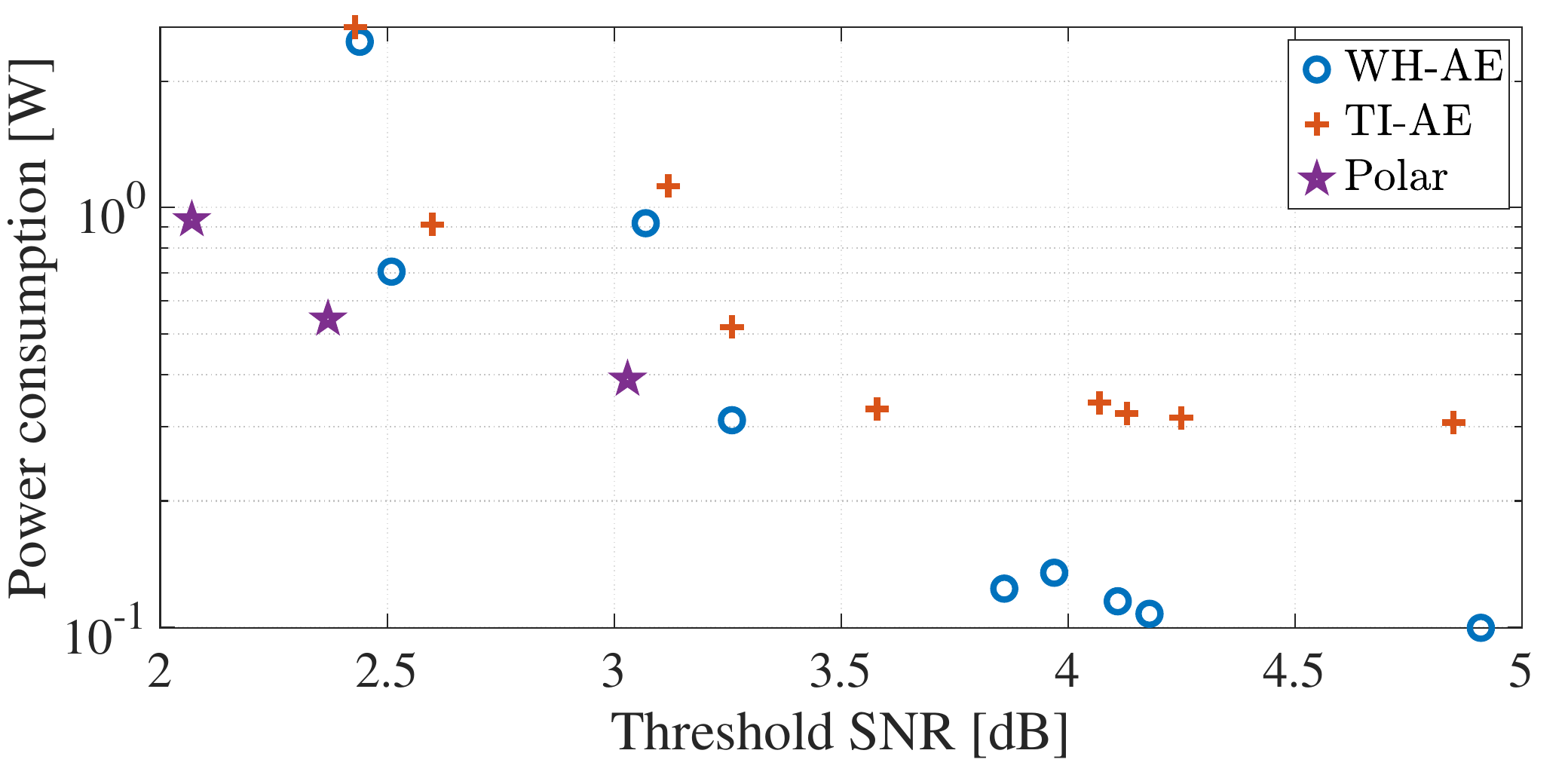}
\caption{System power consumption for coding systems with $n=32$, $R=\tfrac{1}{2}$ and BLER threshold $P_e=0.1\%$. The complexity of each system is scaled by varying the architectural parameters ($ Q$ and $ V$ according to Table~\ref{table:hyperparam} for AE's, and $L=2,4,8$ for the Polar baseline).}
\label{fig:power_snr_single_rate}
\end{figure}

Our first analysis in Fig.~\ref{fig:power_snr_single_rate} focuses on the power consumption of different complexity coding systems with $R=\tfrac{1}{2}$, comparing the coding performance in terms of threshold SNR to achieve $P_e=0.1\%$. The complexity varies according to the architectural parameters $Q$ and $V$ for the AE systems (the number of neurons per layer and the number of layers, respectively), and the parameter $L=2, 4, 8$ for the Polar coding baseline.

As expected, for more complex systems power consumption is higher and coding performance is better (lower threshold SNR). The power consumption of the proposed WH-AE system is consistently lower than that of the baseline TI-AE systems, thanks to its energy-efficient WH-interleaved conversion architecture. Since the Polar system energy efficiency is extrapolated from ASIC synthesis results at $n=256$\cite{ercan_error-correction_2017}, the results shown here should be treated as an order of magnitude estimate, whereas actual implementation results might vary significantly. Nevertheless, we note the proposed WH-AE achieves power consumption figures in the same range as the Polar coding baseline.

Furthermore, the results demonstrate the tradeoff between power consumption and communication performance achievable through training AE systems of varying complexity. We confirm the conclusion of Section~\ref{subsec:hyperparam}: a low-power, reliable WH-AE system can be achieved through careful hyperparameter optimization. Both the power consumption and SNR performance are comparable to the Polar coding baselines. Specifically, the Pareto-optimal WH-AE system in terms of power consumption and threshold SNR, located in the middle left of the figure, achieves a power consumption between Polar decoders with $L=4$ and $L=8$, with a threshold SNR performance only $ 0.14$ dB higher than the Polar ($L=4$) decoder. 

\begin{figure}[htb]
\centering
\includegraphics[width=\linewidth, keepaspectratio]{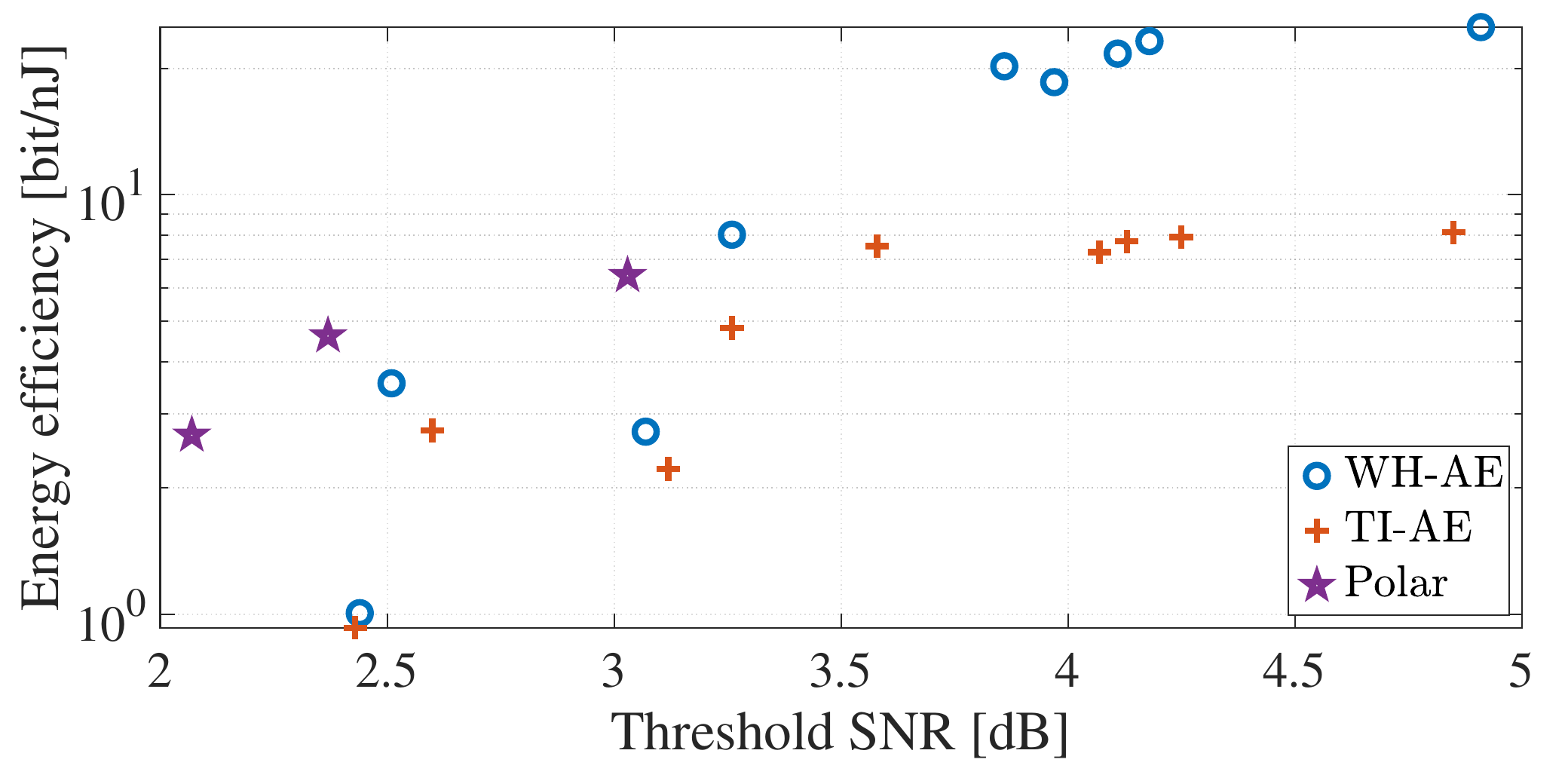}
\caption{System-level energy efficiency for coding systems with $n=32$, $R=\tfrac{1}{2}$ and BLER threshold $P_e=0.1\%$. The complexity of each system is scaled by varying the architectural parameters ($ Q$ and $ V$ according to Table~\ref{table:hyperparam} for AE's, and $L=2,4,8$ for the Polar baseline).}
\label{fig:efficiency_snr_single_rate}
\end{figure}

Starting from the estimated power consumption, we calculate the system-level energy efficiency using \eqref{eq:efficiency}, where higher bit-per-nanoJoule indicates greater efficiency. The results for rate $R=\tfrac{1}{2}$ are shown in Figure~\ref{fig:efficiency_snr_single_rate}, comparing the TI-AE and Polar code baselines with the proposed WH-AE. Once again, the tradeoff between runtime complexity and error correction performance is evident. The optimal WH-AE architecture in a Pareto sense, with minimal threshold SNR and maximum energy efficiency, achieves a threshold SNR of $2.51\,dB$ and an efficiency of $3.55$ bit-per-nJ. The closest baseline is the Polar code with $L=4$, which the proposed WH-AE approaches within $ 0.14$ dB in SNR and 77\% in energy efficiency.

\subsection{Spectral- and energy-efficiency trade-off}\label{subsec:full_tradeoff}
In this section, we explore how the energy efficiency of neural modulation and coding systems scales with the information rate, taking into account their respective information bit throughput. Since conventional coding implementations generally do not report spectral efficiencies $R>1$, we focus only on the AI baseband algorithms and utilize our proposed power consumption model from Section~\ref{sec:powermodel}. Thus, we provide a comprehensive analysis of neural error correction coding systems, examining the tradeoff between information rate, system-level energy efficiency, and coding performance (threshold SNR).

\begin{figure}[!htb]
\includegraphics[width=\linewidth,keepaspectratio]{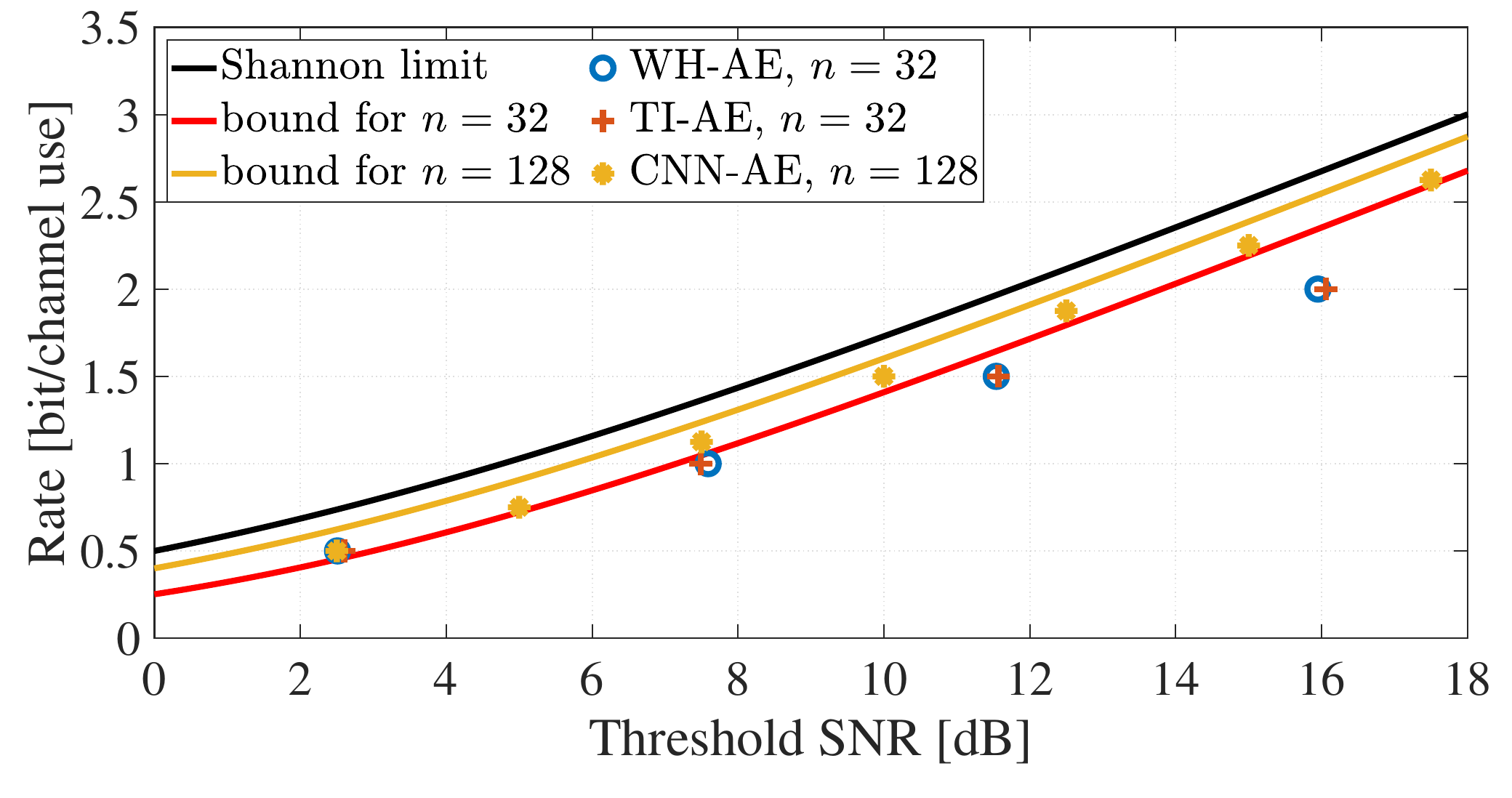}
\caption{Achievable rate versus SNR for BLER target $P_e = 0.1\%$ and block length $n=32$ (the CNN-AE baseline uses $n=128$). The approximate bound for finite block length $n$ is defined in \eqref{eq:rate_bound}.}
\label{fig:rate_snr}
\end{figure}

To assess scaling with higher information rates, the threshold SNR for the proposed method and the neural baselines is shown in Figure~\ref{fig:rate_snr} as a function of the achieved rate $R$. Using the Pareto-optimal hyperparameters as determined in section~\ref{subsec:hyperparam} and \ref{subsec:pareto_hyperparam}, we compare the error correction performance of the proposed WH-AE model to the baseline TI-AE and CNN-AE. The comparison in Figure~\ref{fig:rate_snr} also includes the Shannon \eqref{eq:shannon} and finite block length coding limits \eqref{eq:rate_bound}. 

Fig.~\ref{fig:rate_snr} shows that the proposed WH-AE can achieve error coding performance close to the finite block length upper bound, and especially close to the performance of the TI-AE baseline, which shows that the hardware-friendly WH-AE does not reduce communication performance. Specifically, the coding performance difference between TI-AE and WH-AE is at maximum only $0.11$ dB. Authors of the CNN-AE results\cite{hesham_coding_2023} opted for a larger block length $n=128$ and a less strict BLER threshold $P_e=1\%$; it is thus no surprise that the CNN-AE appears to perform closer to the Shannon limit. Nevertheless, both $n=32$ and $n=128$ AI-based codes perform close to their respective upper bounds. For rate $R=0.5$ the models slightly outperform the finite block length bound for $n=32$, because the bound is not tight for low $n$ and relatively high $P_e$ values~\cite{polyanskiy_channel_2010}.

\begin{figure}[htb]
\centering
\includegraphics[width=\linewidth, keepaspectratio]{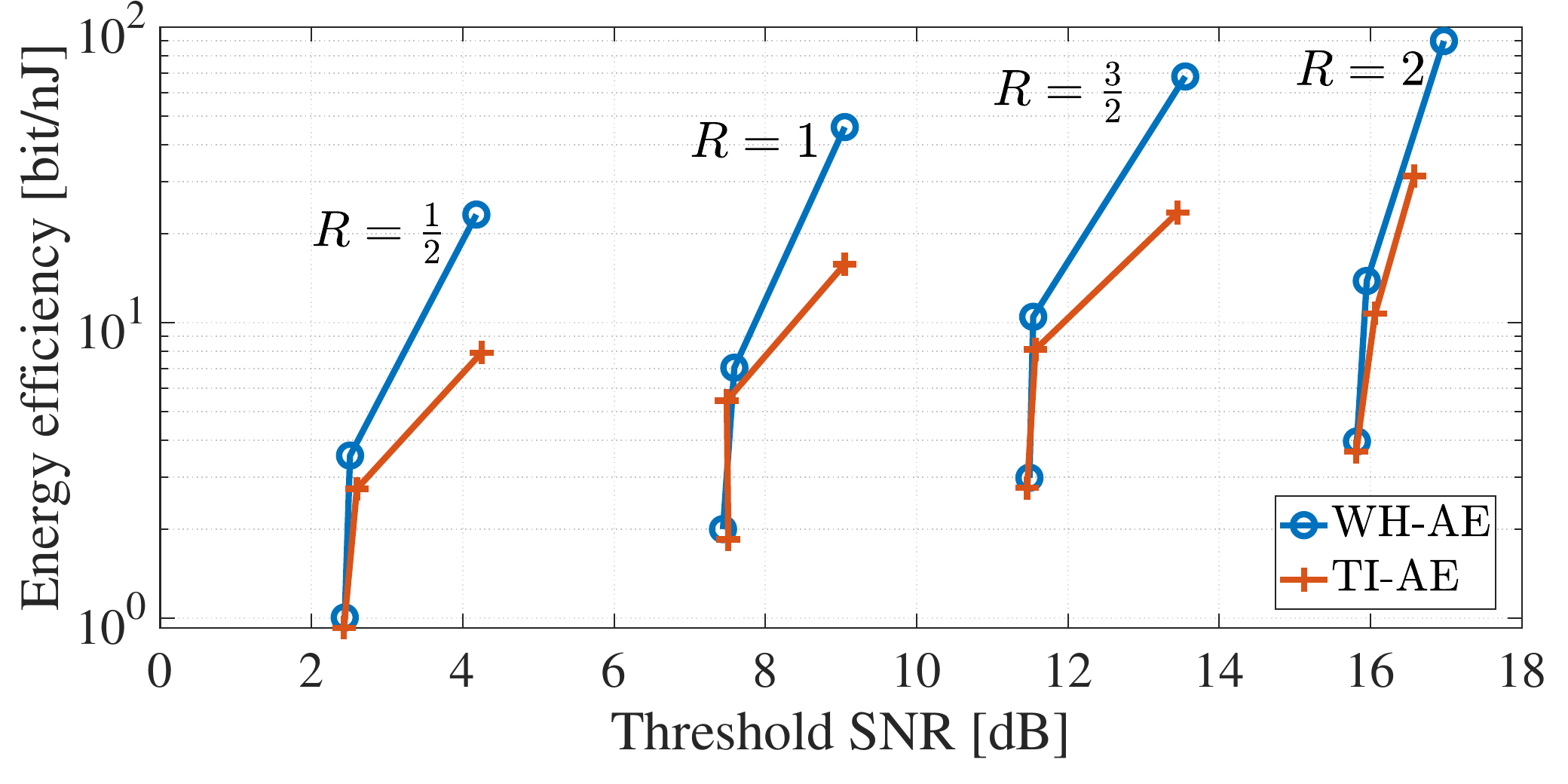}
\caption{System energy efficiency versus threshold SNR for $n=32$ and BLER threshold of $P_e = 0.1\%$. The curves are grouped for rate $R=\tfrac{1}{2}, 1, \tfrac{3}{2}, 2$. The Pareto optimum is the knee point in each curve, transitioning between the most reliable and most energy-efficient models.}
\label{fig:efficiency_snr_pareto}
\vspace{-2mm} % to reduce white space after figure
\end{figure}

Figure~\ref{fig:efficiency_snr_pareto} shows the estimated energy efficiency of Walsh-Hadamard domain autoencoders compared to TI-AE baselines, illustrating the design space spanned by autoencoder models of varying complexity (as controlled by the parameters $Q$ and $V$ in Section~\ref {subsec:hyperparam}). For each spectral efficiency $R$, we show the performance of the least complex, most complex, and Pareto-optimal models. The proposed WH-domain AE models consistently outperform the TI-AE baseline in energy efficiency and threshold SNR to reach a BLER of $10^{-3}$, due to the efficiency of WH-domain conversion.  For the most energy-efficient models, which lie at the top right of each respective curve, the energy efficiency of WH-AE is 2-2.5 times higher than that of TI-AE. For the Pareto-optimal hyperparameters (the middle points of each curve), the WH-AE achieves an average improvement of 29\% in energy efficiency over the TI-AE baseline.

\begin{figure}[htb]
\centering
\includegraphics[width=\linewidth, keepaspectratio]{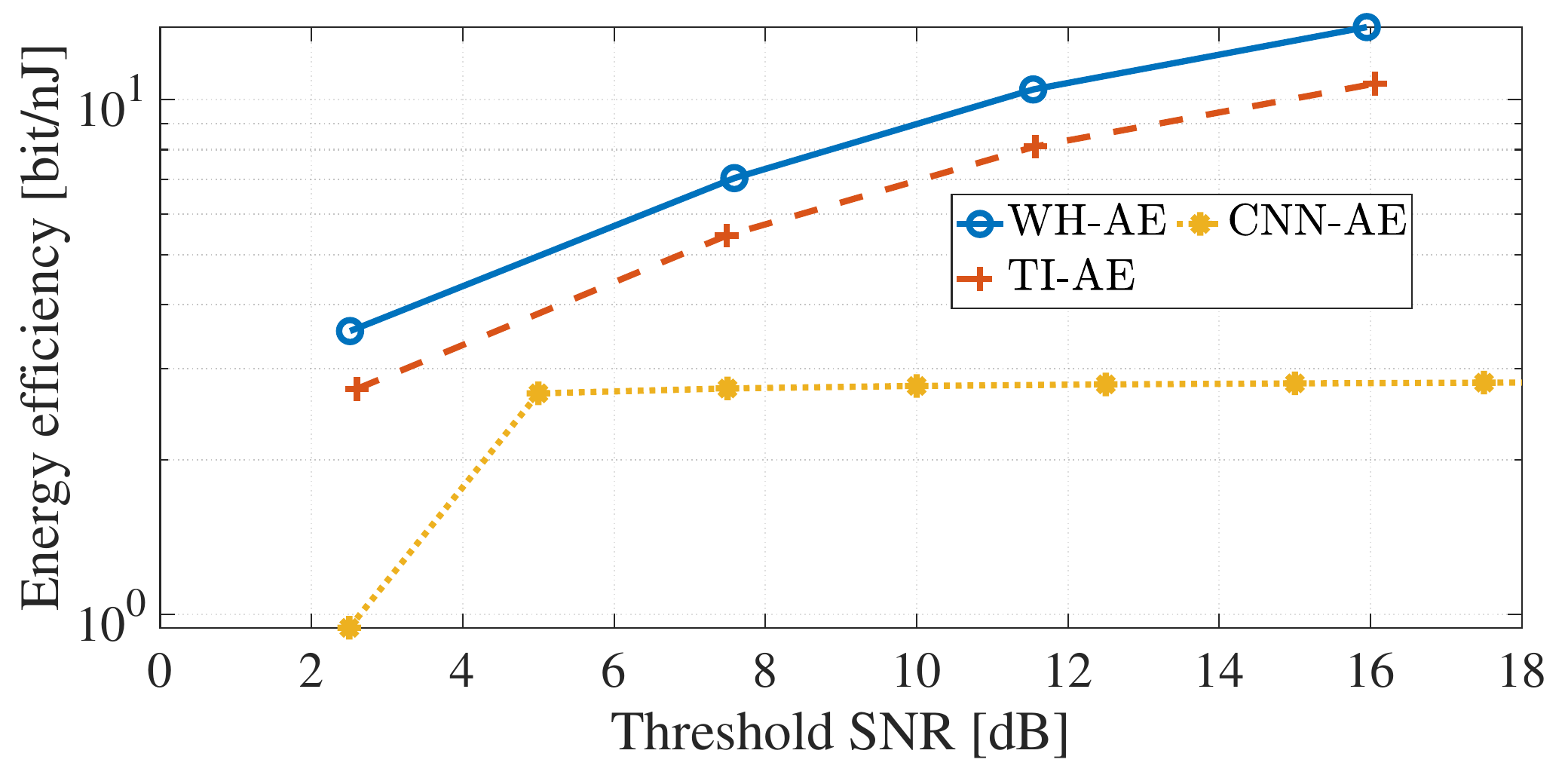}
\caption{System energy efficiency versus threshold SNR for $n=32$ and BLER threshold of $P_e = 0.1\%$ and information rate increasing from $R=0.5$ onwards. The individual points correspond to results in Fig.~\ref{fig:rate_snr}, with both figures sharing the same x-axis.}
\label{fig:energy_efficiency_snr}
\vspace{-2mm} % to reduce white space after figure
\end{figure}

Focusing now on the performance of the Pareto-optimal architecture, as we do not have the full design tradeoff for the CNN-AE baseline, Figure~\ref{fig:energy_efficiency_snr} shows the estimated system-level energy efficiency for the proposed WH-domain transceiver architecture and the neural baselines over different information rates. The individual points correspond to the same models shown in Figure~\ref{fig:rate_snr}. The figure shows the same 29\% improvement in energy efficiency in our proposed WH-AE compared to the TI-AE. Note that the CNN-AE results show an 'outlier' in energy efficiency for $R=\tfrac{1}{2}$; it seems there is no good combination of architectural hyperparameters for that operating point. 

The most noticeable conclusion that can be drawn from the estimated system-level energy efficiency in Fig.~\ref{fig:energy_efficiency_snr} is the difference in scaling between CNN-AE and WH-AE with respect to the information rate (or equivalently, with the threshold SNR). The authors in \cite{hesham_coding_2023} opted for a CNN-AE model that grows in complexity and power consumption with increasing rate. In contrast, our proposed method has nearly constant complexity across different rates, as it uses hidden layers of fixed size. The only scaling with information rate in our proposed architecture is due to the change in $k$, which necessitates a change in the models' input and output dimensions (see Section~\ref{subsec:architecture} for architectural details). As a result, our proposed WH-AE system exhibits nearly constant power consumption with increasing information rate, resulting in energy efficiency that is $1.3$ to $4.8$ times higher than that of the CNN-AE.

\section{Conclusion} \label{sec:conclusion}
This work presents a novel transceiver architecture that integrates Walsh-Hadamard domain interleaved data converters with end-to-end autoencoder-based baseband processing for energy-efficient wideband communications. By leveraging the orthogonality and hardware efficiency of the WH transform, the proposed transceiver enables high data rate communication with improved energy efficiency compared to conventional time-interleaved architectures. Our results demonstrate that E2E autoencoders can be effectively adapted to the WH domain, achieving reliability and throughput comparable to those of state-of-the-art coded modulation schemes while improving energy efficiency.

Extensive hyperparameter optimisation and Monte Carlo simulation results support our conclusion that the WH-domain wideband transceiver achieves competitive error correction performance at short block lengths and offers a favorable tradeoff between runtime complexity and error correction capability. The system-level energy efficiency model, incorporating both wideband converter and intelligent baseband contributions, highlights the energy savings enabled by the WH-domain approach, which amount to a mean improvement of 29 percent over the TI-AE and up to a factor of 4.8 over the CNN-AE. These findings confirm the promise of combining hardware-aware learning with WH interleaving to realize energy-efficient wideband communication systems.

\bibliographystyle{IEEEtran}
\bibliography{references}

@inproceedings{thys_walsh-domain_2024,
	title = {Walsh-domain {Neural} {Network} for {Power} {Amplifier} {Behavioral} {Modelling} and {Digital} {Predistortion}},
	doi = {10.1109/ISCAS58744.2024.10557970},
	urldate = {2024-07-09},
	booktitle = {2024 {IEEE} {International} {Symposium} on {Circuits} and {Systems} ({ISCAS})},
	author = {Thys, Cel and Alonso, Rodney Martinez and Lhomel, Antoine and Fellmann, Maxandre and Deltimple, Nathalie and Rivet, Francois and Pollin, Sofie},
	month = may,
	year = {2024},
}

@article{polyanskiy_channel_2010,
	title = {Channel {Coding} {Rate} in the {Finite} {Blocklength} {Regime}},
	volume = {56},
	issn = {1557-9654},
	doi = {10.1109/TIT.2010.2043769},
	number = {5},
	journal = {IEEE Transactions on Information Theory},
	author = {Polyanskiy, Yury and Poor, H. Vincent and Verdu, Sergio},
	month = may,
	year = {2010},
	pages = {2307--2359},
}

@article{schmidt_data_2020,
	title = {Data {Converter} {Interleaving}: {Current} {Trends} and {Future} {Perspectives}},
	volume = {58},
	issn = {1558-1896},
	shorttitle = {Data {Converter} {Interleaving}},
	doi = {10.1109/MCOM.001.1900683},
	number = {5},
	journal = {IEEE Communications Magazine},
	author = {Schmidt, Christian and Yamazaki, Hiroshi and Raybon, Gregory and Schvan, Peter and Pincemin, Erwan and Yoo, S. J. Ben and Blumenthal, Daniel J. and Mizuno, Takayuki and Elschner, Robert},
	month = may,
	year = {2020},
	pages = {19--25},
}

@article{ferrer_walsh-based_2023,
	title = {A {Walsh}-{Based} {Arbitrary} {Waveform} {Generator} for {5G} {Applications} in 28nm {FD}-{SOI} {CMOS} {Technology}},
	volume = {11},
	issn = {2169-3536},
	doi = {10.1109/ACCESS.2023.3326530},
	urldate = {2024-01-19},
	journal = {IEEE Access},
	author = {Ferrer, Pierre and Rivet, François and Lapuyade, Hervé and Deval, Yann},
	year = {2023},
	pages = {117434--117442},
}

@inproceedings{dehos_d-band_2022,
	title = {D-band channel aggregation receiver architecture based on {IF} analog processing using digital wavelet},
	doi = {10.1109/ICECS202256217.2022.9970798},
	urldate = {2024-04-12},
	booktitle = {2022 29th {IEEE} {International} {Conference} on {Electronics}, {Circuits} and {Systems} ({ICECS})},
	author = {Dehos, Cedric and Gulfo, Jorge-Luis Monsalve and Lachartre, David and Belot, Didier and Courouve, Pierre and Pelissier, Michael},
	month = oct,
	year = {2022},
	pages = {1--4},
}

@misc{adc_survey,
   author = {Murmann, Boris},
   title = {{ADC Performance Survey 1997-2025}},
   note = {[Online]. Available: \url{https://github.com/bmurmann/ADC-survey}}
}

@article{cammerer_trainable_2020,
	title = {Trainable {Communication} {Systems}: {Concepts} and {Prototype}},
	volume = {68},
	issn = {1558-0857},
	shorttitle = {Trainable {Communication} {Systems}},
	doi = {10.1109/TCOMM.2020.3002915},
	number = {9},
	journal = {IEEE Transactions on Communications},
	author = {Cammerer, Sebastian and Aoudia, Fayçal Ait and Dörner, Sebastian and Stark, Maximilian and Hoydis, Jakob and ten Brink, Stephan},
	month = sep,
	year = {2020},
	pages = {5489--5503},
}

@article{manganaro_introduction_2022,
	title = {An {Introduction} to {High} {Sample} {Rate} {Nyquist} {Analog}-to-{Digital} {Converters}},
	volume = {2},
	issn = {2644-1349},
	doi = {10.1109/OJSSCS.2022.3212028},
	journal = {IEEE Open Journal of the Solid-State Circuits Society},
	author = {Manganaro, Gabriele},
	year = {2022},
	pages = {82--102},
}

@inproceedings{ramkaj_33_2019,
	title = {3.3 {A} {5GS}/s 158.{6mW} 12b {Passive}-{Sampling} 8×-{Interleaved} {Hybrid} {ADC} with 9.4 {ENOB} and 160.{5dB} {FoMS} in 28nm {CMOS}},
	doi = {10.1109/ISSCC.2019.8662490},
	urldate = {2025-04-08},
	booktitle = {2019 {IEEE} {International} {Solid}-{State} {Circuits} {Conference} - ({ISSCC})},
	author = {Ramkaj, Athanasios and Ramos, Juan Carlos Pena and Lyu, Yifan and Strackx, Maarten and Marcel Pelgrom, J. M. and Steyaert, Michiel and Verhelst, Marian and Tavernier, Filip},
	month = feb,
	year = {2019},
	pages = {62--64},
}

@article{balasubramanian_systematic_2011,
	title = {Systematic {Analysis} of {Interleaved} {Digital}-to-{Analog} {Converters}},
	volume = {58},
	issn = {1558-3791},
	doi = {10.1109/TCSII.2011.2172526},
	number = {12},
	urldate = {2025-04-29},
	journal = {IEEE Transactions on Circuits and Systems II: Express Briefs},
	author = {Balasubramanian, S. and Creech, G. and Wilson, J. and Yoder, S. M. and McCue, J. J. and Verhelst, M. and Khalil, W.},
	month = dec,
	year = {2011},
	pages = {882--886},
}

@inproceedings{rajapaksha_low_2020,
	title = {Low {Complexity} {Autoencoder} based {End}-to-{End} {Learning} of {Coded} {Communications} {Systems}},
	doi = {10.1109/VTC2020-Spring48590.2020.9128456},
	urldate = {2025-05-12},
	booktitle = {2020 {IEEE} 91st {Vehicular} {Technology} {Conference} ({VTC2020}-{Spring})},
	author = {Rajapaksha, Nuwanthika and Rajatheva, Nandana and Latva-aho, Matti},
	month = may,
	year = {2020},
	pages = {1--7},
}

@article{oshea_introduction_2017,
	title = {An {Introduction} to {Deep} {Learning} for the {Physical} {Layer}},
	volume = {3},
	issn = {2332-7731},
	doi = {10.1109/TCCN.2017.2758370},
	number = {4},
	journal = {IEEE Transactions on Cognitive Communications and Networking},
	author = {O’Shea, Timothy and Hoydis, Jakob},
	month = dec,
	year = {2017},
	pages = {563--575},
}

@inproceedings{hesham_coding_2023,
	title = {Coding for the {Gaussian} {Channel} in the {Finite} {Blocklength} {Regime} {Using} a {CNN}-{Autoencoder}},
	doi = {10.1109/BlackSeaCom58138.2023.10299743},
	urldate = {2024-08-12},
	booktitle = {2023 {IEEE} {International} {Black} {Sea} {Conference} on {Communications} and {Networking} ({BlackSeaCom})},
	author = {Hesham, Nourhan and Bouzid, Mohamed and Abdel-Qader, Ahmad and Chaaban, Anas},
	month = jul,
	year = {2023},
	pages = {15--20},
}

@article{dorner_learning_2023,
	title = {Learning {Joint} {Detection}, {Equalization} and {Decoding} for {Short}-{Packet} {Communications}},
	volume = {71},
	issn = {1558-0857},
	doi = {10.1109/TCOMM.2022.3228648},
	number = {2},
	urldate = {2024-03-12},
	journal = {IEEE Transactions on Communications},
	author = {Dörner, Sebastian and Clausius, Jannis and Cammerer, Sebastian and Brink, Stephan ten},
	month = feb,
	year = {2023},
	pages = {837--850},
}

@article{honkala_deeprx_2021,
	title = {{DeepRx}: {Fully} {Convolutional} {Deep} {Learning} {Receiver}},
	volume = {20},
	issn = {1558-2248},
	shorttitle = {{DeepRx}},
	doi = {10.1109/TWC.2021.3054520},
	number = {6},
	journal = {IEEE Transactions on Wireless Communications},
	author = {Honkala, Mikko and Korpi, Dani and Huttunen, Janne M. J.},
	month = jun,
	year = {2021},
	pages = {3925--3940},
}

@article{marasinghe_waveform_2025,
	title = {Waveform {Learning} {Under} {Phase} {Noise} {Impairment} for {Sub}-{THz} {Communications}},
	volume = {73},
	issn = {1558-0857},
	doi = {10.1109/TCOMM.2024.3430982},
	number = {1},
	urldate = {2025-02-26},
	journal = {IEEE Transactions on Communications},
	author = {Marasinghe, Dileepa and Nguyen, Le Hang and Mohammadi, Jafar and Chen, Yejian and Wild, Thorsten and Rajatheva, Nandana},
	month = jan,
	year = {2025},
	pages = {117--131},
}

@article{ait_aoudia_waveform_2022,
	title = {Waveform {Learning} for {Next}-{Generation} {Wireless} {Communication} {Systems}},
	volume = {70},
	issn = {1558-0857},
	doi = {10.1109/TCOMM.2022.3164060},
	number = {6},
	journal = {IEEE Transactions on Communications},
	author = {Ait Aoudia, Fayçal and Hoydis, Jakob},
	month = jun,
	year = {2022},
	pages = {3804--3817},
}

@INPROCEEDINGS{gruber_dldecoding_2017,
  author={Gruber, Tobias and Cammerer, Sebastian and Hoydis, Jakob and Brink, Stephan ten},
  booktitle={2017 51st Annual Conference on Information Sciences and Systems (CISS)}, 
  title={On deep learning-based channel decoding}, 
  year={2017},
  volume={},
  number={},
  pages={1-6},
  doi={10.1109/CISS.2017.7926071}}

@inproceedings{felix_ofdm-autoencoder_2018,
	title = {{OFDM}-{Autoencoder} for {End}-to-{End} {Learning} of {Communications} {Systems}},
	doi = {10.1109/SPAWC.2018.8445920},
	booktitle = {2018 {IEEE} 19th {International} {Workshop} on {Signal} {Processing} {Advances} in {Wireless} {Communications} ({SPAWC})},
	author = {Felix, Alexander and Cammerer, Sebastian and Dörner, Sebastian and Hoydis, Jakob and Ten Brink, Stephan},
	month = jun,
	year = {2018},
	pages = {1--5},
}

@INPROCEEDINGS{jiang_learncodes_2019,
  author={Jiang, Yihan and Kim, Hyeji and Asnani, Himanshu and Kannan, Sreeram and Oh, Sewoong and Viswanath, Pramod},
  booktitle={ICC 2019 - 2019 IEEE International Conference on Communications (ICC)}, 
  title={LEARN Codes: Inventing Low-Latency Codes via Recurrent Neural Networks}, 
  year={2019},
  volume={},
  number={},
  pages={1-7},
  doi={10.1109/ICC.2019.8761286}}

@ARTICLE{aoudia_modelfree_2019,
  author={Aoudia, Fayçal Ait and Hoydis, Jakob},
  journal={IEEE Journal on Selected Areas in Communications}, 
  title={Model-Free Training of End-to-End Communication Systems}, 
  year={2019},
  volume={37},
  number={11},
  pages={2503-2516},
  doi={10.1109/JSAC.2019.2933891}}

@ARTICLE{buchwald_highspeed_2016,
  author={Buchwald, Aaron},
  journal={IEEE Communications Magazine}, 
  title={High-speed time interleaved ADCs}, 
  year={2016},
  volume={54},
  number={4},
  pages={71-77},
  doi={10.1109/MCOM.2016.7452269}}

@inproceedings{jiang_turbo_2019,
	title = {Turbo {Autoencoder}: {Deep} learning based channel codes for point-to-point communication channels},
	volume = {32},
	shorttitle = {Turbo {Autoencoder}},
	urldate = {2024-03-12},
	booktitle = {Advances in {Neural} {Information} {Processing} {Systems}},
	publisher = {Curran Associates, Inc.},
	author = {Jiang, Yihan and Kim, Hyeji and Asnani, Himanshu and Kannan, Sreeram and Oh, Sewoong and Viswanath, Pramod},
	year = {2019},
}

@article{arikan_channel_2009,
	title = {Channel {Polarization}: {A} {Method} for {Constructing} {Capacity}-{Achieving} {Codes} for {Symmetric} {Binary}-{Input} {Memoryless} {Channels}},
	volume = {55},
	issn = {1557-9654},
	shorttitle = {Channel {Polarization}},
	doi = {10.1109/TIT.2009.2021379},
	number = {7},
	urldate = {2024-06-04},
	journal = {IEEE Transactions on Information Theory},
	author = {Arikan, Erdal},
	month = jul,
	year = {2009},
	pages = {3051--3073},
}

@inproceedings{ercan_error-correction_2017,
	title = {On error-correction performance and implementation of polar code list decoders for {5G}},
	doi = {10.1109/ALLERTON.2017.8262771},
	urldate = {2025-07-18},
	booktitle = {2017 55th {Annual} {Allerton} {Conference} on {Communication}, {Control}, and {Computing} ({Allerton})},
	author = {Ercan, Furkan and Condo, Carlo and Hashemi, Seyyed Ali and Gross, Warren J.},
	month = oct,
	year = {2017},
	pages = {443--449},
}

@inproceedings{kim_communication_2018,
title={Communication Algorithms via Deep Learning},
author={Hyeji Kim and Yihan Jiang and Ranvir B. Rana and Sreeram Kannan and Sewoong Oh and Pramod Viswanath},
booktitle={International Conference on Learning Representations},
year={2018},
url={https://openreview.net/forum?id=ryazCMbR-},
}

@misc{ioffe_batch_2015,
	title = {Batch {Normalization}: {Accelerating} {Deep} {Network} {Training} by {Reducing} {Internal} {Covariate} {Shift}},
	shorttitle = {Batch {Normalization}},
	doi = {10.48550/arXiv.1502.03167},
	urldate = {2025-07-25},
	publisher = {arXiv},
	author = {Ioffe, Sergey and Szegedy, Christian},
	month = mar,
	year = {2015},
	note = {arXiv:1502.03167 [cs]},
}

@misc{hinton_improving_2012,
	title = {Improving neural networks by preventing co-adaptation of feature detectors},
	doi = {10.48550/arXiv.1207.0580},
	urldate = {2025-07-25},
	publisher = {arXiv},
	author = {Hinton, Geoffrey E. and Srivastava, Nitish and Krizhevsky, Alex and Sutskever, Ilya and Salakhutdinov, Ruslan R.},
	month = jul,
	year = {2012},
	note = {arXiv:1207.0580 [cs]},
}

@article{buchwald_high-speed_2016,
	title = {High-speed time interleaved {ADCs}},
	volume = {54},
	issn = {1558-1896},
	doi = {10.1109/MCOM.2016.7452269},
	number = {4},
	urldate = {2024-07-09},
	journal = {IEEE Communications Magazine},
	author = {Buchwald, Aaron},
	month = apr,
	year = {2016},
	pages = {71--77},
}

@inproceedings{bouassida_concurrent_2016,
	title = {A concurrent transmitter in {CMOS} 28nm {FDSOI} technology based on {Walsh} sequences generator},
	doi = {10.1109/NEWCAS.2016.7604779},
	urldate = {2024-02-07},
	booktitle = {2016 14th {IEEE} {International} {New} {Circuits} and {Systems} {Conference} ({NEWCAS})},
	author = {Bouassida, Nassim and Rivet, François and Deval, Yann and Duperray, David and Cathelin, Andreia},
	month = jun,
	year = {2016},
	pages = {1--4},
}

@article{fellmann_block-based_2023,
	title = {A {Block}-{Based} {LMS} {Using} the {Walsh} {Transform} for {Digital} {Predistortion} of {Power} {Amplifiers}},
	volume = {71},
	issn = {1558-0857},
	doi = {10.1109/TCOMM.2023.3294959},
	number = {10},
	urldate = {2024-11-08},
	journal = {IEEE Transactions on Communications},
	author = {Fellmann, Maxandre and Rivet, François and Deltimple, Nathalie and Kerhervé, Eric and Lapuyade, Hervé and Deval, Yann},
	month = oct,
	year = {2023},
	pages = {6074--6087},
}

@article{harmuth_applications_1969,
	title = {Applications of {Walsh} functions in communications},
	volume = {6},
	issn = {1939-9340},
	doi = {10.1109/MSPEC.1969.5214175},
	number = {11},
	journal = {IEEE Spectrum},
	author = {Harmuth, Henning F.},
	month = nov,
	year = {1969},
	pages = {82--91},
}

@misc{sionna2025,
 title = {Sionna},
 author = {Hoydis, Jakob and Cammerer, Sebastian and {Ait Aoudia}, Fayçal and
 Nimier-David, Merlin and Maggi, Lorenzo and Marcus, Guillermo and Vem, Avinash and Keller,
 Alexander},
 note = {https://nvlabs.github.io/sionna/},
 year = {2025},
 version = {1.1.0}
}

@misc{tavildar_polar_2017,
 title = {C and MATLAB implementation for Polar encoding and decoding},
 author = {Saurabh Tavildar},
 note = {https://github.com/tavildar/Polar},
 year = {2017},
}

@techreport{3gpp.38.214,
 author = {3GPP},
 number = {38.214},
 title = {{NR; Physical layer procedures for data}},
 type = {Technical Specification (TS)},
 url = {https://portal.3gpp.org/desktopmodules/Specifications/SpecificationDetails.aspx?specificationId=3216},
 year = {2017}
}

@article{chakraborty_twirld_2024,
	title = {{TWIRLD}: {Transformer} {Generated} {Terahertz} {Waveform} for {Improved} {Radio} {Link} {Distance}},
	issn = {2831-316X},
	shorttitle = {{TWIRLD}},
	doi = {10.1109/TMLCN.2024.3483111},
	urldate = {2024-10-21},
	journal = {IEEE Transactions on Machine Learning in Communications and Networking},
	author = {Chakraborty, Shuvam and Parisi, Claire and Saha, Dola and Thawdar, Ngwe},
	year = {2024},
	pages = {1--1},
}

@inproceedings{cui_6g_2025,
	title = {{6G} {Wireless} {Communications} in 7–24 {GHz} {Band}: {Opportunities}, {Techniques}, and {Challenges}},
	shorttitle = {{6G} {Wireless} {Communications} in 7–24 {GHz} {Band}},
	doi = {10.1109/DySPAN64764.2025.11115934},
	urldate = {2025-09-26},
	booktitle = {2025 {IEEE} {International} {Symposium} on {Dynamic} {Spectrum} {Access} {Networks} ({DySPAN})},
	author = {Cui, Zhuangzhuang and Zhang, Peize and Pollin, Sofie},
	month = may,
	year = {2025},
	pages = {1--8},
}

@article{elfikky_symbol_2024,
	title = {Symbol {Detection} and {Channel} {Estimation} for {Space} {Optical} {Communications} {Using} {Neural} {Network} and {Autoencoder}},
	volume = {2},
	issn = {2831-316X},
	doi = {10.1109/TMLCN.2023.3346811},
	urldate = {2025-08-26},
	journal = {IEEE Transactions on Machine Learning in Communications and Networking},
	author = {Elfikky, Abdelrahman and Rezki, Zouheir},
	year = {2024},
	pages = {110--128},
}

@misc{wiesmayr_design_2024,
	title = {Design of a {Standard}-{Compliant} {Real}-{Time} {Neural} {Receiver} for {5G} {NR}},
	doi = {10.48550/arXiv.2409.02912},
	urldate = {2025-08-29},
	publisher = {arXiv},
	author = {Wiesmayr, Reinhard and Cammerer, Sebastian and Aoudia, Fayçal Aït and Hoydis, Jakob and Zakrzewski, Jakub and Keller, Alexander},
	month = sep,
	year = {2024},
	note = {arXiv:2409.02912 [cs]},
}

@article{bioglio_design_2021,
	title = {Design of {Polar} {Codes} in {5G} {New} {Radio}},
	volume = {23},
	issn = {1553-877X},
	doi = {10.1109/COMST.2020.2967127},
	number = {1},
	urldate = {2024-01-30},
	journal = {IEEE Communications Surveys \& Tutorials},
	author = {Bioglio, Valerio and Condo, Carlo and Land, Ingmar},
	year = {2021},
	pages = {29--40},
}

@article{kneip_impact_2023,
	title = {{IMPACT}: {A} 1-to-4b 813-{TOPS}/{W} 22-nm {FD}-{SOI} {Compute}-in-{Memory} {CNN} {Accelerator} {Featuring} a 4.2-{POPS}/{W} 146-{TOPS}/mm2 {CIM}-{SRAM} {With} {Multi}-{Bit} {Analog} {Batch}-{Normalization}},
	volume = {58},
	issn = {1558-173X},
	shorttitle = {{IMPACT}},
	doi = {10.1109/JSSC.2023.3269098},
	number = {7},
	urldate = {2025-08-29},
	journal = {IEEE Journal of Solid-State Circuits},
	author = {Kneip, Adrian and Lefebvre, Martin and Verecken, Julien and Bol, David},
	month = jul,
	year = {2023},
	pages = {1871--1884},
}

@article{fino_unified_1976,
	title = {Unified {Matrix} {Treatment} of the {Fast} {Walsh}-{Hadamard} {Transform}},
	volume = {C-25},
	issn = {1557-9956},
	doi = {10.1109/TC.1976.1674569},
	number = {11},
	urldate = {2025-09-17},
	journal = {IEEE Transactions on Computers},
	author = {{Fino} and {Algazi}},
	month = nov,
	year = {1976},
	pages = {1142--1146},
}

@article{jornet_evolution_2024,
	title = {The {Evolution} of {Applications}, {Hardware} {Design}, and {Channel} {Modeling} for {Terahertz} ({THz}) {Band} {Communications} and {Sensing}: {Ready} for {6G}?},
	issn = {1558-2256},
	shorttitle = {The {Evolution} of {Applications}, {Hardware} {Design}, and {Channel} {Modeling} for {Terahertz} ({THz}) {Band} {Communications} and {Sensing}},
	doi = {10.1109/JPROC.2024.3412828},
	urldate = {2025-10-03},
	journal = {Proceedings of the IEEE},
	author = {Jornet, Josep M. and Petrov, Vitaly and Wang, Hua and Popović, Zoya and Shakya, Dipankar and Siles, Jose V. and Rappaport, Theodore S.},
	year = {2024},
	pages = {1--32},
}

@article{ferrer_experimental_2025,
	title = {Experimental {Demonstration} of {Walsh}-{Based} {RF} {Conversion} {Using} {Wideband} {DAC} and {ADC} in 28 nm {FDSOI} {CMOS} {Technology}},
	issn = {1558-0806},
	doi = {10.1109/TCSI.2025.3615682},
	urldate = {2025-10-30},
	journal = {IEEE Transactions on Circuits and Systems I: Regular Papers},
	author = {Ferrer, Pierre and Fellmann, Maxandre and Rivet, François and Deltimple, Nathalie and Lapuyade, Hervé and Kerhervé, Eric and Deval, Yann},
	year = {2025},
	pages = {1--9},
}

@ARTICLE{hoydis_2021_6gainative,

  author={Hoydis, Jakob and Aoudia, Fayçal Ait and Valcarce, Alvaro and Viswanathan, Harish},

  journal={IEEE Communications Magazine}, 

  title={Toward a 6G AI-Native Air Interface}, 

  year={2021},

  volume={59},

  number={5},

  pages={76-81},

  doi={10.1109/MCOM.001.2001187}}

@article{radulov_28-nm_2015,
	title = {A 28-nm {CMOS} 7-{GS}/s 6-bit {DAC} {With} {DfT} {Clock} and {Memory} {Reaching} {SFDR} {\textgreater}50 {dB} {Up} to 1 {GHz}},
	volume = {23},
	issn = {1557-9999},
	doi = {10.1109/TVLSI.2014.2350540},
	number = {9},
	urldate = {2026-01-09},
	journal = {IEEE Transactions on Very Large Scale Integration (VLSI) Systems},
	author = {Radulov, Georgi I. and Quinn, Patrick J. and van Roermund, Arthur H. M.},
	month = sep,
	year = {2015},
	pages = {1941--1945},
}

@ARTICLE{tal_list_2015,

  author={Tal, Ido and Vardy, Alexander},

  journal={IEEE Transactions on Information Theory}, 

  title={List Decoding of Polar Codes}, 

  year={2015},

  volume={61},

  number={5},

  pages={2213-2226},

  doi={10.1109/TIT.2015.2410251}}

\end{document}